%% file: MAINradVP.tex
\documentclass[11pt,3p]{elsarticle}

\usepackage{slashed}
\usepackage{graphicx}
\usepackage{amssymb,amstext,amsbsy,amsmath,amsfonts}
\usepackage{nicefrac}
\usepackage{multirow}

\usepackage{xcolor}

\usepackage{lineno}

\usepackage{hyperref}

\usepackage{upgreek}

\biboptions{sort&compress}

\DeclareMathOperator\arctanh{arctanh}

\newcommand{\<}{\langle}
\renewcommand{\>}{\rangle}
\newcommand{\txts}{\textstyle}
\newcommand{\beq}{\begin{equation}}
\newcommand{\eeq}{\end{equation}}
\newcommand{\bea}{\begin{eqnarray}}
\newcommand{\eea}{\end{eqnarray}}
\newcommand{\nn}{\nonumber}
\def\eqlab#1{\label{eq:#1}}

\def\barr{\left(\begin{array}{c}}
\def\earr{\end{array}\right)}
\def\bmat{\left(\begin{array}{cc}}
\def\emat{\end{array}\right)}
\def\eref#1{(\ref{eq:#1})}
\def\Eqref#1{Eq.~(\ref{eq:#1})}

\def\Figref#1{Fig.~\ref{fig:#1}}

\DeclareMathOperator{\im}{Im}
\def\nn{\nonumber}

\def\MM{\mathcal{M}}

\def\al{\alpha}

\def\veps{\varepsilon}  

\def\la{\lambda}

\def\si{\sigma}

\begin{document}

\begin{frontmatter}

  \begin{flushright}
    {MITP-22-082}
  \end{flushright}
  \medskip
  
 
 \title{Forward light-by-light scattering \\ and electromagnetic correction  to hadronic vacuum polarization}

\author[label 1]{Volodymyr~Biloshytskyi}
\author[label 1]{En-Hung~Chao}
\author[label 5]{Antoine~G\'erardin}
\author[label 3]{Jeremy~R.\ Green}
\author[label 1,label 2]{Franziska~Hagelstein}
\author[label 1,label 4]{Harvey~B.\ Meyer}
\author[label 1]{Julian~Parrino}
\author[label 1]{Vladimir~Pascalutsa}

\address[label 1]{PRISMA$^+$ Cluster of Excellence \& Institut f\"ur Kernphysik, \\ Johannes Gutenberg Universit\"at Mainz,  D-55128 Mainz, Germany}
\address[label 3]{School of Mathematics and Hamilton Mathematics Institute, Trinity College, Dublin 2, Ireland}
\address[label 2]{Paul Scherrer Institut, CH-5232 Villigen PSI, Switzerland}
\address[label 4]{Helmholtz Institut Mainz, Staudingerweg 18, D-55128 Mainz, Germany}
\address[label 5]{Aix Marseille Univ., Universit\'e de Toulon, CNRS, CPT, Marseille, France}

\begin{abstract}
Lattice	QCD calculations of the hadronic vacuum polarization (HVP) have reached a precision where the electromagnetic (e.m.) correction can no longer be neglected. This correction is both computationally challenging and hard to validate, as it leads to ultraviolet (UV) divergences and to sizeable infrared (IR) effects associated with the massless photon.
While we precisely determine the UV divergence using the operator-product expansion, we propose to introduce a separation scale $\Lambda\sim400\;$MeV into the internal photon propagator, whereby the calculation splits into a short-distance part, regulated in the UV by the lattice and in the IR by the scale $\Lambda$, and a UV-finite long-distance part to be treated with coordinate-space methods, thereby avoiding power-law finite-size effects altogether.
In order to predict the long-distance part, we express the UV-regulated e.m.\ correction to the HVP via the forward hadronic light-by-light (HLbL) scattering amplitude and relate the latter via a dispersive sum rule to $\gamma^*\gamma^*$ fusion cross-sections. Having tested the relation by reproducing the two-loop QED vacuum polarization (VP) from the tree-level $\gamma^*\gamma^*\to e^+e^-$ cross-section, we predict the   expected lattice-QCD integrand resulting from the $\gamma^*\gamma^*\to\pi^0$ process.
\end{abstract}


\end{frontmatter}

\section{Introduction}

\begin{figure}[hb]
\centering
\includegraphics[width=0.22\linewidth]{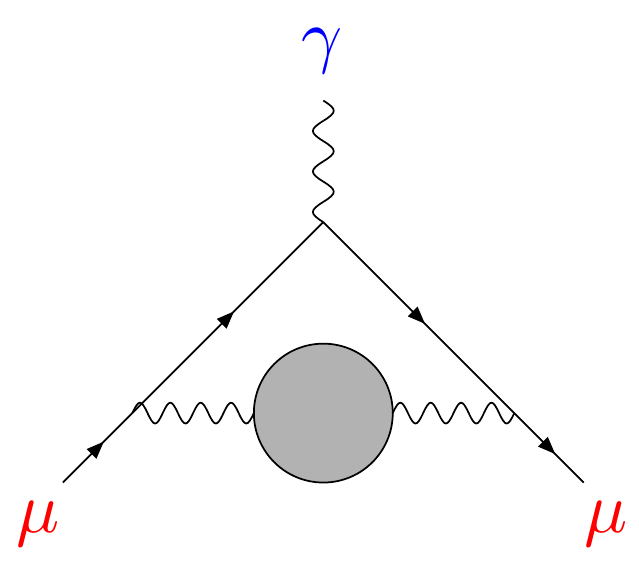}
\hspace{0.5cm}
\includegraphics[width=0.22\linewidth]{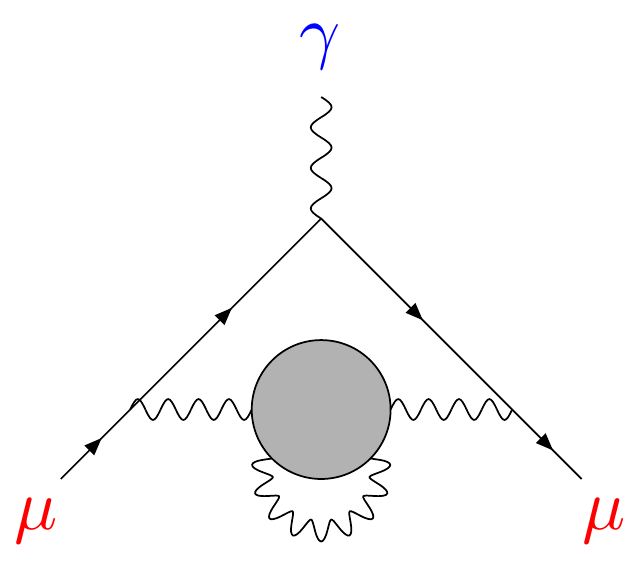}
  \caption{Different hadronic  contributions to $(g-2)_\mu$: leading-order HVP (left), e.m.\ corrections to leading-order HVP (right). The blob represents the contribution from QCD alone.} 
  \label{fig:HVPdiagrams}
\end{figure}

The long-standing discrepancy between theory~\cite{Aoyama:2012wk,Aoyama:2019ryr,Czarnecki:2002nt,Gnendiger:2013pva,Davier:2017zfy,Keshavarzi:2018mgv,Colangelo:2018mtw,Hoferichter:2019mqg,Davier:2019can,Keshavarzi:2019abf,Kurz:2014wya,Melnikov:2003xd,Masjuan:2017tvw,Colangelo:2017fiz,Hoferichter:2018kwz,Gerardin:2019vio,Bijnens:2019ghy,Colangelo:2019uex,Blum:2019ugy,Colangelo:2014qya} and experiment~\cite{Bennett:2006fi,Muong-2:2021ojo} for the muon $g-2$ has recently been challenged by several precision lattice QCD calculations of the HVP contribution (cf.\ Fig.~\ref{fig:HVPdiagrams}) from intermediate hadronic distance scales~\cite{Blum:2018mom,Borsanyi:2020mff,Ce:2022kxy,Alexandrou:2022amy,FermilabLattice:2022izv,Colangelo:2022vok}. One of the lattice-QCD based calculations 
has already reached a subpercent-level of precision for the full leading-order HVP contribution~\cite{Borsanyi:2020mff}, thus becoming competitive with the data-driven dispersive method, which has traditionally been used to evaluate this contribution.
At this level of precision, care must also be taken of the leading isospin-breaking corrections, both the strong isospin-breaking effect stemming from the unequal $u$ and $d$ quark masses,
and the e.m.\ effect arising from the quarks carrying electric charges, as shown by the second diagram in Fig.~\ref{fig:HVPdiagrams}.
These effects are taken into account by lattice collaborations (e.g., Ref.~\cite{Blum:2018mom,Borsanyi:2020mff,Risch:2021hty}); however, few stringent cross-checks are possible at present. First of all, these effects depend on the precise point in the parameter space of isospin-symmetric QCD, which is not exactly the same in different calculations. Furthermore, it has not been possible to rigorously compare the size of these effects to phenomenological predictions, partly because the (QED) radiative correction to the HVP is divergent if one does not account for the counterterms associated with the quark masses and the strong-coupling, whose finite parts depend on the conventional choice of the `physical point' in isospin-symmetric QCD~\cite{deDivitiis:2013xla}.

Here we propose a computational strategy that enables much more direct comparisons between lattice QCD and hadron phenomenology.
In its simplest incarnation, the idea is to add and subtract a Pauli-Villars term~\cite{Pauli:1949zm} to the photon propagator\footnote{Such a decomposition of the photon propagator has been found to be helpful in other contexts; see in particular Ref.\ \cite{Sirlin:1980nh}.},
\beq\eqlab{gampropsplit} 
\frac{1}{k^2} = \left(\frac{1}{k^2} - \frac{1}{k^2+\Lambda^2}\right) + \frac{1}{k^2+\Lambda^2},
\eeq
where $k$ is the Euclidean four-momentum and $\Lambda$ is a typical hadronic scale. 
The first term leads to a UV-finite effect on the HVP and is sensitive to long-distance contributions such as the $\pi^0\gamma$ and $\eta\gamma$ channels; it can be treated analogously to the HLbL contribution to $g-2$ by using coordinate-space methods~\cite{Asmussen:2016lse,Blum:2017cer,Chao:2021tvp}, whereby
power-law effects due to the internal photon propagators are avoided.
The second can be treated entirely in lattice regularization by having the photon field defined on the same lattice as the QCD fields~\cite{deDivitiis:2013xla}. However, since a photon mass is now present, no issue with the photon zero-mode arises, nor do power-law finite-size effects occur. We return to this aspect in section~\ref{sec:lattstrat} but 
 remark here that a number of different methods have been used in the extensive literature
on incorporating the coupling of quarks to photons into lattice QCD calculations
(see \cite{Blum:2010ym,Ishikawa:2012ix,Aoki:2012st,Borsanyi:2014jba,Endres:2015gda,Horsley:2015eaa,Fodor:2016bgu,Giusti:2017dmp,Boyle:2017gzv,Feng:2021zek,Frezzotti:2022dwn,Portelli:2015wna,Patella:2017fgk,FlavourLatticeAveragingGroupFLAG:2021npn} for a representative set of publications).

\begin{figure}[thb]
\centering
	\includegraphics[width=0.5\textwidth]{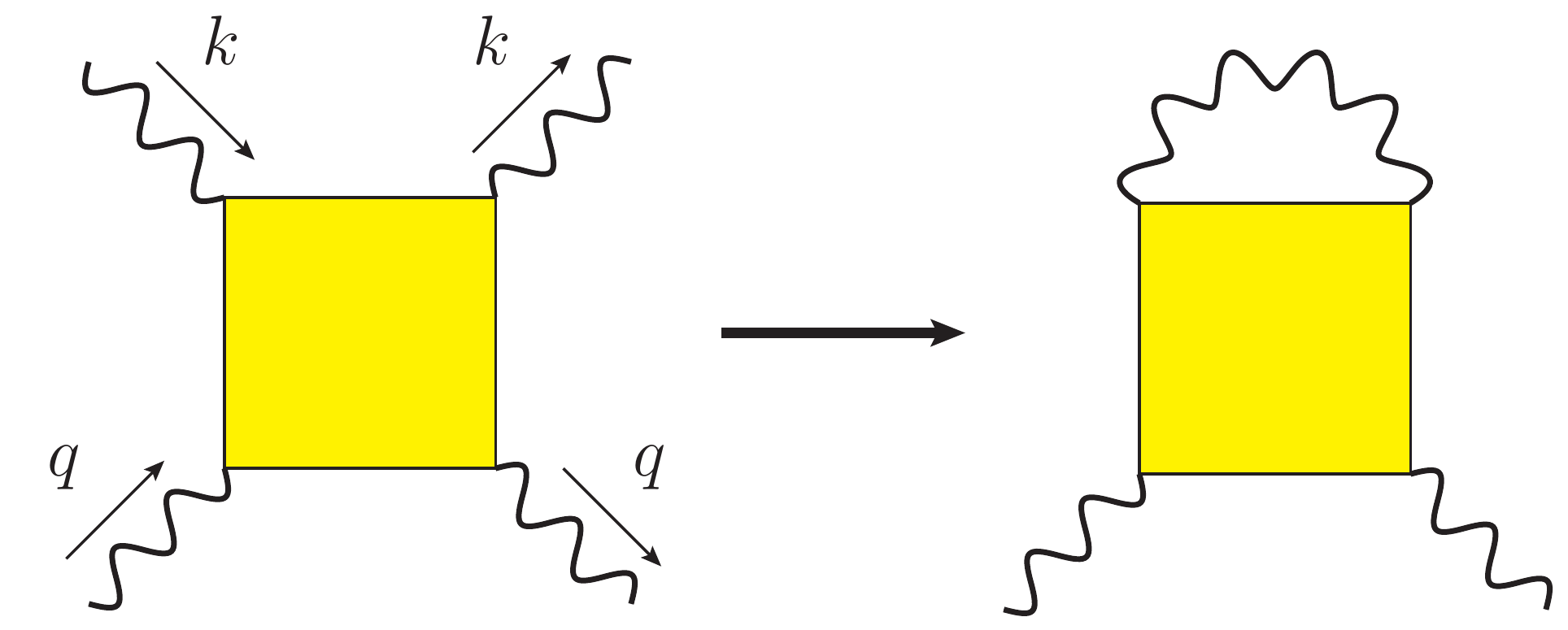}
	\caption{Cottingham-like formula for QCD LbL amplitude.}
	\label{fig:QCDblob}
\end{figure}

How then can one predict the leading QED correction to the HVP with a UV-regularized photon propagator in place? We shall express it through the forward HLbL amplitude~\cite{Knecht:2003kc,Blokland:2001pb,Pascalutsa:2017ta}, as
shown in Fig.~\ref{fig:QCDblob}. As has been noted~\cite{Pascalutsa:2017ta}, the connection between the forward HLbL amplitude and the e.m.\ correction to the HVP bears a strong resemblance with the Cottingham formula~\cite{Cottingham:1963zz,Walker-Loud:2012ift,Gasser:2020mzy,Gasser:2020hzn,Gasser:2015dwa}, which expresses e.m.\ mass splittings in terms of the forward Compton scattering amplitude. The analogy becomes apparent if one views light-by-light (LbL) scattering as Compton scattering off a photon.

However, not much practical use has so far emerged from this connection.
Perhaps the main reason is that the insertion of two standard $\bar q q \gamma$ vertices leads to a divergence, requiring the insertion of O($e^2$) counterterms to cancel it.
From the standpoint of the HLbL amplitude, the divergence appears due to the forward HLbL falling off too slowly (as $1/k^2$) for one of the incoming photon momenta becoming large.
However, the first term on the right-hand side of \Eqref{gampropsplit} amounts to a UV-regularization of the photon propagator, and in this case
the integral over the forward HLbL amplitude yielding an e.m.\ correction to the HVP becomes finite. Therefore, with sufficient knowledge of the forward HLbL amplitude, obtained either by using the dispersive sum rules~\cite{Pascalutsa:2012pr} or direct lattice calculations \cite{Green:2015sra}, one can make a definite prediction for this correction. The comparison can even be done in a  more differential way, at the integrand level, as we shall illustrate.  Also, accumulated knowledge on the HLbL amplitudes, for example concerning the relative importance of the different quark Wick-contraction topologies~\cite{Bijnens:2016hgx,Gerardin:2017ryf}, can be usefully applied to the leading QED corrections to the HVP.

The rest of this paper is organized as follows.
We start in the continuum, deriving in section~\ref{sec:derivn} the relation between the forward HLbL amplitude and the e.m.\ correction to the HVP. In section~\ref{sec:qed}, this relation is tested against known results in a pure QED setting. Section~\ref{sec:ope} contains a derivation of the divergence that develops when the UV-cutoff $\Lambda$ is sent to infinity and indicates in which flavor combinations the divergence partly cancels.
In section~\ref{sect:pi0}, we then come to the prediction for the $\pi^0$ exchange in the forward HLbL amplitude, and thereby to its contribution in the e.m.\ correction to the HVP and ultimately in the muon $(g-2)$.
We then formulate a computational strategy in section~\ref{sec:lattstrat} for computing the leading isospin-breaking effects to the HVP in lattice QCD. The section also discusses aspects of the coordinate-space method and presents the integrand corresponding to the $\pi^0$ exchange contribution.
Finally, section~\ref{sec:concl} summarizes our findings and 
offers an outlook into further possible applications of this work.

\section{A Cottingham-like formula for the radiative correction to the HVP \label{sec:derivn}}

The first correction $\Delta{\Pi}(Q^2)$ to the leading HVP\footnote{In our notation throughout this paper, the HVP contains an additional factor $e^2$ relative to the notation widely used in lattice QCD calculations, for instance in Refs.\ \cite{Blum:2002ii,Bernecker:2011gh}.} $\Pi_{e^2}(Q^2)$ can be written in the form 
\beq \eqlab{Pie4tot}
\Delta{\Pi}(Q^2) = \lim_{\Lambda\to\infty} \Big({\Pi}_{{\rm 4pt}}(Q^2,\Lambda) +{\Pi}_{{\rm ct}}(Q^2,\Lambda)\Big),
\eeq 
where $\Lambda$ is a UV-regularization parameter.
We begin by establishing a formula for $\Pi_{{\rm 4pt}}(Q^2,\Lambda)$, named in this way because it involves the four-point function of the e.m.\ current, which at the same time provides the quantum field-theoretic definition of the LbL scattering amplitude.
The second term, ${\Pi}_{{\rm ct}}(Q^2,\Lambda)$, consists of the required counterterms and the strong-isospin breaking contribution. 
While its precise form is not needed here, more details will be given in section \ref{sec:lattstrat}.

The LbL scattering amplitude $\MM^{\mu_1\mu_2\mu_3\mu_4}$ depends
on the four-momenta of the incoming $(q_1,q_2)$ and outgoing $(q_3,q_4)$ photons. The forward kinematics correspond to $q_1=q_3\equiv k$ and $q_2=q_4\equiv q$, 
see Fig.~\ref{fig:QCDblob}. Contracting the photon-line 1 with 3, we obtain a contribution to the
VP tensor:
\beq 
\eqlab{F0}
\Pi_{{\rm 4pt}}^{\mu_2 \mu_4} (q^2,\Lambda) = \frac{1}{2}
\int \frac{d^4 k}{(2\pi)^4  }\left[ \frac{-i g_{\mu_1\mu_3}}{k^2+i0^+} \right]_\Lambda \MM^{\mu_1\mu_2\mu_3\mu_4}(k, q)\,,
\eeq 
where the factor of one-half is the symmetry factor; in square brackets is the Feynman-gauge photon propagator, regulated at the scale $\Lambda$, for instance \`a la Pauli-Villars, $[1/k^2]_{\Lambda}= 1/k^2 - 1/(k^2-\Lambda^2)$.
Due to gauge invariance, the VP tensor has the following general form, 
\beq 
\Pi^{\mu\nu}(q) = \Pi(q^2) (q^2 g^{\mu\nu} - q^\mu q^\nu ) ,
\eeq 
and hence its scalar part can be expressed as:  
\beq 
\eqlab{F1}
\Pi_{{\rm 4pt}}(q^2,\Lambda)= \frac{1}{6q^2} 
\int \frac{d^4 k}{(2\pi)^4} \left[\frac{-i}{k^2+i0^+}\right]_\Lambda \MM(k, q)\,,
\eeq 
where 
\beq \eqlab{calMdef}
\MM\equiv g_{\mu_1\mu_3}g_{\mu_2\mu_4}\MM^{\mu_1\mu_2\mu_3\mu_4}(k,q)\,,
\eeq 
is the traced LbL amplitude. 
The latter is a scalar function of three invariants:  $k^2$, $q^2$, and $\nu\equiv k\cdot q$. It is even in $\nu$ and symmetric under the interchange of $k$ and $q$. We shall write it as $\MM(\nu, K^2, Q^2)$, where $K^2=-k^2$ and $Q^2=-q^2$ will further be assumed to be positive, i.e., the photons are spacelike. 

Introducing the helicity LbL amplitudes as 
\beq 
M_{\la_1\la_2\la_3\la_4} = \veps_{\la_1}^{\mu_1} (q_1)\,\veps_{\la_2}^{\mu_2} (q_2)\,\veps_{\la_3}^{\ast\mu_3}(q_3) \veps_{\la_4}^{\ast\mu_4}(q_4)
\, \MM_{\mu_1\mu_2\mu_3\mu_4}\,,
\eeq 
 with $\veps_\la^\mu (q)$ the photon
 polarization vectors,  the traced amplitude  can be written as \cite{Budnev:1971sz}:
 \beq 
 \MM =  \sum_{\la,\si=\pm,0} (-1)^{\la+\si}
M_{\la\si\la\si} =  4 \MM_{TT} -  2 \MM_{LT} - 2\MM_{TL} +  
\MM_{LL}\,,
 \eeq 
 where  
\bea 
&& \MM_{TT} =  \mbox{$\frac12$} \big(M_{++++}+M_{+-+-}\big), \quad \MM_{LL} = M_{0000}\,,\nn\\
&& \MM_{LT} = M_{0+0+}, \quad \MM_{TL} = M_{+0+0}\,.
\eea  
For spacelike photon virtualities, the optical theorem relates the imaginary part of these amplitudes to a $\gamma^*\gamma^*$-fusion cross section \cite[Eq.~(16)]{Pascalutsa:2012pr},
so that
\beq
\im \MM(\nu,k^2,q^2) =2 \sqrt{X}\, \sigma(\nu,k^2,q^2),
\eeq
where $\si = 4\si_{TT}-2\si_{TL}-2\si_{LT}+\si_{LL}$, and $X=\nu^{2}-q^2\,k^2$. 
Furthermore,
the analytic properties of the $\nu$-dependence warrant a dispersive
representation.
Since all relevant LbL amplitudes are even in $\nu$ and require one subtraction \cite[Section II.\ C]{Pascalutsa:2012pr}, the dispersion relation takes the  form
\bea
\MM(\nu ,K^2,\,Q^2) &=& \MM(\bar\nu ,K^2,\,Q^2)+\overline{\MM}(\nu ,K^2,\,Q^2),\nn\\
\overline{\MM}(\nu ,K^2,\,Q^2) &=& \frac{2}{\pi} (\nu^2-\bar\nu^2)\int\limits_{\nu_\mathrm{thr.}}^\infty d \nu'\,
\frac{ \nu' \im \MM(\nu', K^2,Q^2)}{(\nu^{\prime\,2}-\bar\nu^2)(\nu^{\prime\,2}-\nu^2)},
\eqlab{Mdisprep}
\eea
where we are free to choose any subtraction point $ \bar\nu$, and
$\nu_\mathrm{thr.}$ is the lowest particle production threshold. For example, in QED,   $\nu_\mathrm{thr.} = \nicefrac12 (K^2+Q^2)+ 2m_e^2$ is the threshold for $e^+e^-$ production; see \ref{sec:CrossSectionsDef} for further details.

The dispersive representation justifies the Wick rotation in the
evaluation of \Eqref{F1} and we obtain the Cottingham formula analogue:
\bea 
\Pi_{{\rm 4pt}}(Q^2,\Lambda) &=& \frac{1}{6Q^4(2\pi)^3} \int\limits_0^\infty d K^2  \left[\frac{1}{K^2}\right]_\Lambda\, \int\limits_{0}^{K^2Q^2} d\nu^2 \, \left(\frac{K^2Q^2}{\nu^2}-1 \right)^{1/2}\!
\MM(\nu ,\,K^2,\,Q^2)\,.
\eqlab{CottinghamFormulaFinal}
\eea
We refer the reader to \Eqref{DRform} in appendix for the dispersive form, which, up to one subtraction, expresses
this contribution in terms of $\gamma^*\gamma^*$-fusion cross sections.
As indicated in \Eqref{Pie4tot}, this contribution must be combined with the appropriate counterterm, to which we return in section~\ref{sec:lattstrat}, in order to obtain the first correction $\Delta\Pi(Q^2)$ to the HVP.

At this point, let us briefly comment on the flavor structure of the HLbL amplitude $\MM$, particularly regarding to isospin, which plays an important role at low energies.
The e.m.\ current carried by the quarks contains 
both an isovector and an isoscalar component. 
The LbL amplitude can be written as the sum of the three partial contributions where  (i) all four currents are isovector; (ii) all four currents are isoscalar; and (iii) in one pair of currents, both are isovector, while in the complementary pair, both are isoscalar, and one sums over all six possible pairings.
Pole contributions of isovector mesons such as the pion only occur in the third contribution, while isoscalar-meson exchanges appear in all three contributions.

\section{Reproducing the two-loop QED vacuum polarization \label{sec:qed}}

\begin{figure}[hb]
\centering
	\includegraphics[width=\textwidth]{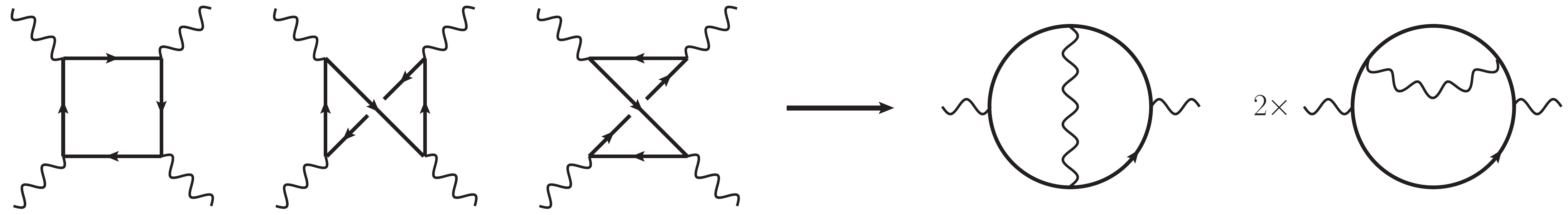}
	\caption{One-loop LbL scattering (left three diagrams) and the resulting two-loop VP.}
	\label{fig:TwoLoopQED}
\end{figure}

In order to test our Cottingham analogue, \Eqref{CottinghamFormulaFinal}, 
we apply it to the QED VP: 
we expect the one-loop LbL amplitude to provide the 
two-loop VP, see Fig.~\ref{fig:TwoLoopQED}.

Substituting the one-loop LbL amplitude (cf.\ \ref{sec:CrossSectionsDef}) into \Eqref{CottinghamFormulaFinal} yields a complicated expression, which
we only show here in the expanded form:
\bea
\Pi_{{\rm 4pt}}(Q^2,\Lambda) 
&=& \frac{\alpha^2}{\pi^2}
\left[
-\frac{1}{2}
+\frac{329}{1620} \frac{Q^2}{m_\ell^2}
-\frac{2333}{75600}\left(\frac{Q^2}{m_\ell^2}\right)^2
+\frac{43579}{7938000}\left(\frac{Q^2}{m_\ell^2}\right)^3
+O(Q^8)
\right]\nn\\ &-&\frac{\alpha^2}{\pi^2}\log\frac{\Lambda}{m_\ell}
\left[-\frac{1}{2}+
\frac{1}{5}\frac{Q^2}{m_\ell^2}
-\frac{3}{70}\left(\frac{Q^2}{m_\ell^2}\right)^2
+\frac{1}{105}\left(\frac{Q^2}{m_\ell^2}\right)^3
+O(Q^8)
\right] ,
\eea
up to terms that vanish for ${\Lambda}\to\infty$. Hereafter $m_\ell$ stands for the lepton mass appearing in the loops. In this calculation we have in fact adopted a simpler, momentum-cutoff regularization:
$[\frac{1}{k^2}]_\Lambda = \frac{\theta(\Lambda^2-k^2)}{k^2}$. In the present context, this form of regularization is equivalent to Pauli-Villars regularization, up to terms suppressed by $1/\Lambda^2$.

The counterterm, $\overline{\Pi}_{{\rm ct}}(Q^2,\Lambda)$, can be obtained by applying the standard rules of renormalized perturbation theory, for which we use  the Pauli-Villars regularization, see \ref{sec:qedctpv}: 
\bea
\overline{\Pi}_{{\rm ct}}(Q^2,\Lambda) &=& 6\,\frac{\alpha^2}{\pi^2}\left(\frac{1}{4}+\log\frac{\Lambda}{m_\ell}\right)\Bigg[
\frac{1}{6}
-\frac{1}{\kappa^2}
+\frac{4}{\kappa^3\sqrt{4+\kappa^2}}
\arctanh{\left(\frac{\kappa}{\sqrt{4+\kappa^2}}\right)}
\Bigg] \nn\\&=& \frac{\alpha^2}{\pi^2}\left(\frac{1}{4}+\log\frac{\Lambda}{m_\ell}\right)
\left[
\frac{1}{5}\kappa^2
-\frac{3}{70}\kappa^4
+\frac{1}{105}\kappa^6
+O(\kappa^8)
\right],
\label{NonlocalCounterterm}
\eea
where $\kappa =  Q/m_\ell$.
Altogether [cf.\ \Eqref{Pie4tot}], we obtain the following result for the small-$Q^2$ expansion of the two-loop VP:
\beq\label{eq:VPbarexp}
\Delta\overline{\Pi}(Q^2) = 
\frac{\alpha^2}{\pi^2}
\left[
\frac{41}{162}\frac{Q^2}{m_\ell^2}
-\frac{449}{10800}\left(\frac{Q^2}{m_\ell^2}\right)^2
+\frac{62479}{7938000}\left(\frac{Q^2}{m_\ell^2}\right)^3
+O(Q^8)
\right].
\eeq
Note that:
\beq\overline\Pi(Q^2) \equiv \Pi(Q^2) - \Pi(0),
\eeq
with $\Pi_{{\rm 4pt}}(0,\Lambda)$ given in \Eqref{Pi0subtraction}.
To check this result, we use the well-known dispersion relation: 
\beq
\overline{\Pi}(Q^2) = -\frac{Q^2}{\pi}\int_{4m_{\ell}^2}^{\infty}\;
\frac{d t}{t(t+Q^2)}\im \Pi(t).
\label{DispersiveAnswer}
\eeq
together with the well-known  O($e^4$) imaginary part, first computed in 1955~\cite{Kallen:1955fb} as well as in~\cite{Lautrup:1968tdb}, given for instance in~\cite{SchwingerTextBook}
(Eq.~(5-4.200) on p.~109) and in~\cite{Charles:2017snx}. 
Using this well-known result, we reproduce Eq.~\eqref{eq:VPbarexp} obtained via the Cottingham formula.
We have also checked this numerically for arbitrary $Q^2$.

\subsection{Evaluation of the fourth-order vacuum polarization contribution to the muon $(g-2)$}
 
We now test the Cottingham-like formula beyond the small-$Q^2$ expansion and compute the fourth-order VP contribution to the muon $(g-2)$ using numerical integration.
In order to avoid  numerical instabilities emerging in loop integrals, that in the case of the virtual LbL amplitude we found out to persist in all familiar packages for one-loop numerical integration, we choose the more elegant way and calculate the LbL amplitude $\MM(\nu ,K^2,\,Q^2)$ itself via the dispersive approach; see \Eqref{Mdisprep}.

The imaginary part $\im  \MM(\nu, K^2,Q^2)$ at the tree level in QED,
as well as expressions for the subtraction term $\MM(\bar\nu ,K^2,\,Q^2)$  for the two choices of subtraction point $\bar\nu_1=0$ and $\bar\nu_2=KQ$ are provided in~\ref{sec:CrossSectionsDef}.

\begin{figure}[t]
    \centering
    \includegraphics[width=0.7\textwidth]{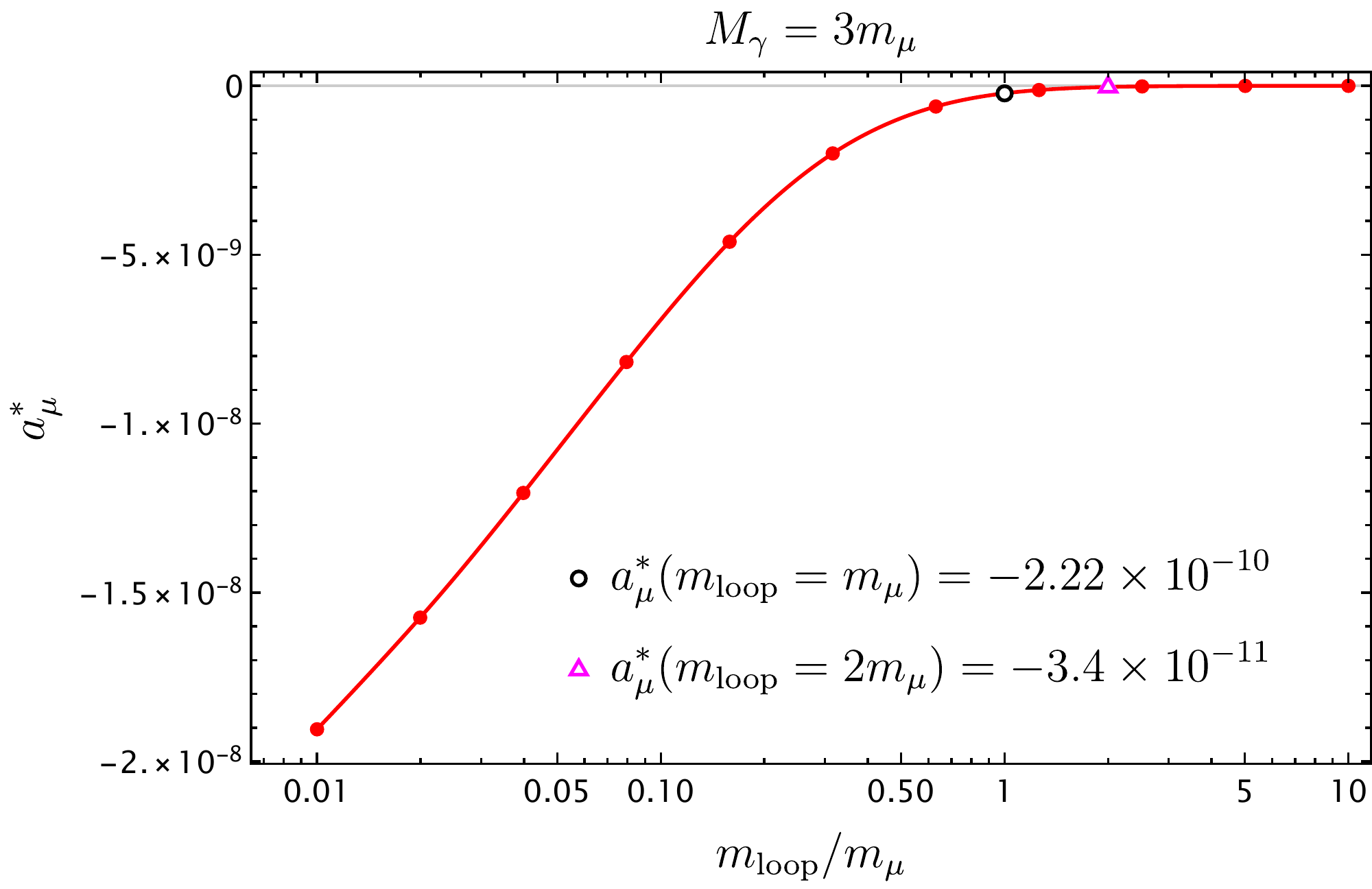}
    \caption{Anomalous magnetic moment via the Cottingham formula without the counterterm to be compared with lattice QCD: varying the mass inside the VP. The cutoff scale is chosen to be $\Lambda\equiv M_\gamma = 3m_\mu$.}
    \label{fig:a_mu_ast}
\end{figure}

We extracted the `Thomson limit' $\Pi_{{\rm 4pt}}(0,\Lambda)$  from the first term in the series expansion of $\MM$, and after performing the integration over $x$, we arrive at \Eqref{Pi0subtraction}.
In particular, this quantity is finite for a fixed value of $\Lambda$.
From here, the O($e^6$) contribution of the regulated fourth-order QED VP to the anomalous magnetic moment can  be computed using the general formula \cite{Lautrup:1969nc,Lautrup:1971jf,deRafael:1993za}:
\beq 
a_\upmu^{\mathrm{VP}} =  \frac{\al}{\pi}\int\limits_0^\infty d Q^2 \, \mathcal{K}(Q^2)\, \overline \Pi(Q^2),
\eqlab{g2VP}
\eeq 
where the kernel function is given by
\beq 
\mathcal{K}(Q^2) = \frac{1}{2m_\mu^2} \frac{(v-1)^3}{2v(v+1)}, \qquad v = \sqrt{1+\frac{4m_\mu^2}{Q^2}},
\eeq 
with $m_\mu$ the muon mass. The integral over the kernel function alone
 gives the Schwinger term, $\Delta a_\upmu=\al/2\pi$.  Therefore,
substituting the renormalized VP, we obtain:
\beq 
a^{\mathrm{VP}}_\upmu = -\frac{\al}{2\pi}\Pi(0) + \frac{\al}{\pi}\int\limits_0^\infty d Q^2 \, \mathcal{K}(Q^2)\, \Pi(Q^2).
\eqlab{Schwinger}
\eeq 
This means that the subtraction term in $\overline{\Pi}_{{\rm 4pt}}(Q^2,\Lambda)$ yields the contribution 
$-\frac{\al}{2\pi}\Pi_{{\rm 4pt}}(0,\Lambda) $ to $a^{\rm VP}_\upmu$.
Results obtained 
with a Pauli-Villars regulator and $\Lambda=3m_\upmu$ are provided as a function of the lepton mass $m_\ell$ appearing in the VP in Fig.~\ref{fig:a_mu_ast}. Furthermore, adding the contribution of the 
appropriate counterterm and subsequently taking the limit $\Lambda\to\infty$, we obtain the full O($e^4$) QED VP to $a_\upmu$.
For a muon in the VP loop, we obtain:
\begin{align}
   \Delta a_\upmu \simeq
    6.6\times 10^{-10}.\quad
\end{align}
The result numerically agrees with Ref.~\cite{MignacoRemiddi1969}.

\input{OPE}

\input{hvpnlopi0sec}

\input{lattstrat}

\input{coordspace}

\section{Conclusion \label{sec:concl}}

We have written a Cottingham-like formula for the leading QED correction to the HVP, mainly in terms of the traced forward HLbL scattering amplitude.
While the latter is a physical amplitude, when contracting an incoming with an outgoing photon line, the corresponding momentum integral diverges logarithmically in the UV. 
The required counterterms that remove that divergence have been worked out in section \ref{sec:ope} with the help of the OPE: they involve the derivatives with respect to the quark masses and the virtuality of the leading HVP. 
The finite part of the counterterms, however, depends on the precise choice of the point in the parameter-space of isospin-symmetric QCD around which the isospin-breaking effects are computed.

At present, it appears that the most promising application of the Cottingham-like formula is to implement it with a finite regulator on the order of a few hundred MeV.
The regularization amounts to replacing the internal photon line by a Pauli-Villars regulated propagator, or any other convenient form of propagator regularization.
The leading QED correction to the HVP with 
such a regularized photon propagator in place can be computed in lattice QCD  using coordinate-space techniques similar to the calculation of the HLbL contribution to the muon $(g-2)$, without incurring power-law finite-size effects.
When working on very large lattices, as realized in master-field
simulations~\cite{Luscher:2017cjh,Francis:2019muy}, these
techniques are particularly natural~\cite{Ce:2021akh}.
Irrespective of whether one uses this new approach or the more established method involving the removal of the spatial zero-mode of the photon,
a direct comparison  becomes possible between the lattice-QCD calculation and the prediction based on the Cottingham-like formula.
For the latter, we recall that the traced forward HLbL amplitude can be represented dispersively in terms of the $\gamma^*\gamma^*\to {\rm hadrons}$ fusion cross-section, up to one subtraction term. 
The complementary part, i.e. the second term in \Eqref{gampropsplit}, involves a massive photon propagator in the case of the Pauli-Villars regularization choice. For this part, the lattice provides a natural UV-regularization, and a prescription for handling the photon zero mode on a finite lattice is then no longer needed, owing to the photon mass.

Finally, in the context of the original Cottingham formula, it could  be interesting to compute the e.m.\ contribution to the proton-neutron mass difference with a regularized photon propagator, which could then be compared in detail and without scheme uncertainty to predictions based on the (rather mature) dispersive treatment of the forward Compton amplitude (see~\cite{Gasser:2020hzn} and Refs.\ therein).

\section*{Acknowledgements}
We acknowledge the support of  Deutsche Forschungsgemeinschaft (DFG) 
through the research unit FOR 5327 ``Photon-photon interactions in the Standard Model and beyond - exploiting the discovery potential from MESA to the LHC'' (grant 458854507), and as part of the Cluster of Excellence “Precision Physics, Fundamental Interactions and Structure of Matter” (PRISMA+ EXC 2118/1) funded by the DFG within the German Excellence strategy (Project ID 39083149).
F.H. acknowledges the support of the Swiss National Science Foundation (SNSF) through the Ambizione Grant PZ00P2\_193383 and the support of DFG 
through the Emmy Noether Programme under grant 449369623.
H.B.M. and E.-H.C.'s work was supported by the European Research Council (ERC) under the European Union’s Horizon 2020 research and innovation program through Grant Agreement No.\ 771971-SIMDAMA.
J.R.G. acknowledges support from the Simons Foundation through the Simons Bridge for Postdoctoral Fellowships scheme.
\appendix

\input{PolarizedCrossSections}
\input{apdxQED2}

\input{QEDcounterterm}

\input{ccs_kernel}

\bibliographystyle{elsarticle-num-names}
\bibliography{refs.bib}

\end{document}

%% file: OPE.tex
\section{Forward LbL amplitude 
at large virtuality from the Operator Product Expansion \label{sec:ope}}
 
Since the forward HLbL amplitude ${\cal M}(k,q)$
is finite, the divergence of \Eqref{F1} as $\Lambda\to\infty$ can only arise from performing the $k$ integral.
The question is then, what is the large-$k$ behaviour of ${\cal M}(k,q)$ for fixed $q$.
This is a typical application for the Operator Product Expansion (OPE). In this section we work in Euclidean space
and our starting point is 
\bea\eqlab{4ptEucl}
\Big\<V^{\rm em}_\mu(x) V^{\rm em}_\nu(y) V^{\rm em}_\sigma(z) V^{\rm em}_\lambda(0)\Big\>
= \int_{q_1,q_2,q_3} e^{i(q_1x+q_2y+q_3z)}\; \Pi_{\mu\nu\sigma\lambda}(q_1,q_2,q_3),
\eea
 the e.m.\ current carried by the quarks, in units of the positron charge, being given by
$
V^{\rm em}_\mu = \frac{2}{3}\bar u\gamma_\mu u - \frac{1}{3} \bar d \gamma_\mu d -...$
We recall that the HLbL amplitude is directly related to
$\Pi_{\mu\nu\sigma\lambda}$~\cite{Green:2015sra}. In particular,
for the forward amplitude \Eqref{calMdef},
the connection reads
\beq\eqlab{forwEucl}
{\cal M}(k\cdot q, k^2, q^2) = e^4 \delta_{\mu\nu}\delta_{\sigma\lambda}
\Pi_{\mu\nu\sigma\lambda}(-k, k, -q),
\eeq
where the scalar products on the left-hand side are Euclidean.
The large momentum $k$ ``forces" the two vertices $x$
and $y$ to come close together. From a power-counting perspective,
it is the dimension-four operators that can cause a logarithmically
divergent behaviour in \Eqref{F1}, since they
contribute as $O^{(4)}/k^2$. It is then only necessary
to know their Wilson coefficients to order $\alpha_s$ included,
since $\alpha_s(k^2)^2 O^{(4)}/k^2$ multiplied by a photon propagator already yields a UV-finite integral.

Note that the two indices of the vector currents are contracted
with each other (thus cancelling the axial current contribution in the OPE), and that we may average the result over the 
direction of $k$, given that we are interested in subsequently 
integrating over $k$ in \Eqref{F1}.
The result of the OPE can then only contain  operators
with vacuum quantum numbers. 

From a different perspective, the divergence resulting from the integral over the photon momentum $k$ must be removable by
the available counterterms of the theory. Moreover,
since vector currents do not renormalize in QCD, 
the relevant counterterms are only those associated with the parameters of the theory, which are the gauge coupling and 
the quark masses. These parameters are respectively associated
with the operators $G_{\alpha\beta}^aG_{\alpha\beta}^a$
and $m_f\bar\psi_f\psi_f$ for each quark flavor $f$.

The Wilson coefficients of scalar operators appearing in the OPE of QCD currents have been calculated a long time ago~\cite{Chetyrkin:1985kn}. We report only the result up to the order required for our purposes\footnote{As we shall see explicitly in the following subsection, the leading term $(6/k^2) m_f\bar\psi_f\psi_f$ also applies to the QED case. The other displayed terms are for the gauge group SU(3) and do not depend on the number of quark flavors $n_f$. The leading coefficient of $m_f\bar\psi_f \psi_f$ and the leading coefficient of $G_{\alpha\beta}^aG_{\alpha\beta}^a$ are consistent with the calculation of~\cite{Hill:2016bjv}.},
\bea
&& \Big\<\int d^4x\; e^{ikx} \;{\rm T}(\bar\psi_f(x)\gamma_\mu\psi_f(x)\;\bar\psi_f(0)\gamma_\mu\psi_f(0))
\Big\>_{\hat k}
\\ && \stackrel{k^2\to\infty}{=} \frac{3}{k^2}\Big[ 2\Big(1+\frac{\alpha_s}{3\pi}\Big)m_f\bar\psi_f\psi_f
+ \frac{\alpha_s}{12\pi} \Big(1+ \frac{7}{6}\frac{\alpha_s}{\pi}\Big)G_{\alpha\beta}^aG_{\alpha\beta}^a\Big]
\nn\,.
\eea
Interpreting the gluonic operator in terms of the (renormalization group invariant) trace anomaly $\theta(x) = \frac{2\beta(g)}{g} {\cal L}_{\rm g}$, where ${\cal L}_{\rm g}=\frac{1}{4} G_{\alpha\beta}^aG_{\alpha\beta}^a$ is the gluonic Lagrangian density
and $\beta(g) = \mu\frac{\partial g}{\partial \mu} = -g^3(b_0+ b_1 g^2+...)$ the QCD beta function\footnote{In these conventions,
$b_0 = \frac{1}{(4\pi)^2}(11 - \frac{2}{3}n_f)$ and 
$b_1 = \frac{1}{(4\pi)^4}(102 - \frac{38}{3}n_f)$.},
we rewrite
\beq
\frac{\alpha_s}{12\pi} \Big(1+ \frac{7}{6}\frac{\alpha_s}{\pi}\Big)G_{\alpha\beta}^a(x)G_{\alpha\beta}^a(x)
= \frac{-1}{24\pi^2b_0}\,\Big(1+g^2\big(\textstyle{\frac{7}{24\pi^2}-\frac{b_1}{b_0}}\big)+{\rm O}(g^4)\Big)\; \theta(x)\,.
\eeq
Thereby we arrive at the following prediction for
 the asymptotic large-$k^2$ behaviour of the four-point amplitude 
\bea\eqlab{OPE1}
&&  \int \frac{d\Omega_k}{2\pi^2}  \Big\< \int d^4x \int d^4y \;e^{ik(x-y)}\, V^{\rm em}_\mu(x) V^{\rm em}_\mu(y)   V^{(1)}_\sigma(z) V^{(2)}_\lambda(0)\Big\>
\\ &&     \stackrel{k^2\to\infty}{=}
      \frac{3}{k^2} \sum_f {\cal Q}_f^2\Big[ 2 \Big(1+\frac{\alpha_s}{3\pi}\Big) m_f \Big\< \int d^4x\; \bar\psi_f\psi_f V^{(1)}_\sigma(z) V^{(2)}_\lambda(0) \Big\>
 \nn\\ &&  \qquad   -\frac{1}{24\pi^2b_0}\,\Big(1+g^2\big(\textstyle{\frac{7}{24\pi^2}-\frac{b_1}{b_0}}\big)\Big)
      \Big\<\int d^4x\; \theta(x) V^{(1)}_\sigma(z) V^{(2)}_\lambda(0)\Big\>
      \Big]
    \nonumber\,,
\eea
where ${\cal Q}_f =\{\nicefrac23,\,-\nicefrac13,\ldots\}$ are the quark electric charges. At this point, we keep the currents $V^{(1)}_\sigma(z)$ and $V^{(2)}_\lambda(0)$ unspecified, in particular in their flavour structure.

The effect of inserting the mass operator into a correlation function is to differentiate the latter with respect to the quark mass,
\beq
\< A \; \int d^4x\; m\bar\psi_x \psi_x \> = - m \frac{\partial }{\partial m} \<A\>, 
\eeq
while the effect of the trace anomaly on a renormalization-group invariant correlation function of mass-dimension $n$ is to differentiate with respect to all  scales on which the correlation function depends,
\beq
\Big\< A(y,z,...,m_1,m_2,...) \; \int d^4x\; \theta(x) \Big\> 
= \Big(-n -y_\nu\frac{\partial}{\partial y_\nu} 
- z_\nu \frac{\partial}{\partial z_\nu} - \dots
+ \sum_j m_j\frac{\partial}{\partial m_j}\Big) \Big\< A(y,z,...)\Big\>,
\eeq
where $y$ and $z$ are space-time coordinates.
We now set $V^{(1)}_\sigma(z)= V^{\rm em}_\sigma(z)$
and $V^{(2)}_\lambda(0)=V^{\rm em}_\lambda(0)$.
Let $H_{\lambda\sigma}(z)$ be the kernel yielding the leading-order subtracted VP when integrated over with the correlator $e^2\<V^{\rm em}_\sigma(z) V^{\rm em}_\lambda(0)\>$ (see Eqs.\ \ref{eq:CCS_LO} and \ref{eq:CCSmaster} below).
Acting with the linear operator
 \beq \eqlab{getHVP}
 -\frac{e^4}{2} \int \frac{d^4k}{(2\pi)^4}\, \Big[\frac{1}{k^2}\Big]_\Lambda \int d^4z\,H_{\lambda\sigma}(z)\,,
 \eeq
on both sides of \Eqref{OPE1},
we conclude that the asymptotic large-$\Lambda$ behaviour of the four-point amplitude contribution to the fourth-order VP is given by
\bea\eqlab{OPEfinal}
&&   \overline\Pi_{{\rm 4pt}}(Q^2,\Lambda)
     \stackrel{\Lambda\to\infty}{=}
 \frac{3e^2}{8\pi^2 } 
       \sum_f {\cal Q}_f^2\Big[  \Big(
       \log\Big(\frac{\Lambda}{\mu_{\rm IR}}\Big)+\frac{1}{24\pi^2 b_0} \log\Big(\frac{\alpha_s(\mu_{\rm IR})}{\alpha_s(\Lambda)}\Big)\Big) m_f \frac{\partial}{\partial m_f} 
 \\ &&  +\frac{1}{48\pi^2b_0}\,\Big(\log\Big(\frac{\Lambda}{\mu_{\rm IR}}\Big)+ \frac{1}{2b_0} \big(\textstyle{\frac{7}{24\pi^2}-\frac{b_1}{b_0}}\big) \log\Big(\frac{\alpha_s(\mu_{\rm IR})}{\alpha_s(\Lambda)}\Big) \Big) \Big( 2q^2 \frac{\partial}{\partial q^2}
      + \sum_{f'} m_{f'} \frac{\partial}{\partial m_{f'}}\Big)
      \Big]
   \overline\Pi_{e^2}(Q^2).
    \nonumber
\eea

\subsection{Explicit OPE calculation at leading order}

Consider then the OPE of two vector currents at leading order,
\beq
\bar\psi_x \gamma_\mu \psi_x \; \bar \psi_y \gamma_\nu \psi_y
= \bar\psi_x \gamma_\mu S(x-y) \gamma_\nu \psi_y + \bar\psi_y \gamma_\nu S(y-x) \gamma_\mu \psi_x,
\eeq
with $S(x)$ the position-space fermion propagator.
Eventually one finds 
\beq
\int \frac{d\Omega_k}{2\pi^2}\int d^4y\; e^{ik(x-y)}\, \bar\psi_x \gamma_{\mu} \psi_x \; \bar \psi_y \gamma_{\mu} \psi_y 
= \frac{6m}{k^2}  \bar\psi_x \psi_x +  \frac{2}{k^2} \Big(
{\textstyle\frac{1}{2}}\bar\psi_x  \gamma_\alpha (\overrightarrow\partial_\alpha-\overleftarrow{\partial_\alpha})  \psi_x
+m \bar\psi_x \psi_x\Big).
\eqlab{VVlo}
\eeq
We have already noted the effect of the mass-operator insertion
in terms of differentiating the correlation function with respect
to the quark mass.
To understand the effect of inserting  the `equation of motion'
(EOM)
operator appearing in brackets in \Eqref{VVlo},
imagine multiplying the Euclidean quark action by $\lambda$, $S_E(\lambda) = \lambda \bar\psi (D+m)\psi$.
In the Euclidean path integral, we can take expectation values $\<A\>_\lambda$ using $\exp(-S_E(\lambda))$ as weight:
it simply means that each quark propagator contains an additional $1/\lambda$ factor.
Thus, if computing $\<A\>_\lambda$ involves $n_p$ propagators,
\beq
- \frac{\partial}{\partial \lambda} \<A\>_\lambda \Big|_{\lambda=1} = n_p\; \<A\>_{\lambda=1}.
\eeq
On the other hand, the same derivative can be expressed as
\beq
- \frac{\partial}{\partial \lambda} \<A\>_\lambda \Big|_{\lambda=1} =
\< A\; \int d^4x\,
({\textstyle\frac{1}{2}} \bar\psi_x  \gamma_\alpha (\overrightarrow\partial_\alpha-\overleftarrow{\partial_\alpha})  \psi_x +m \bar\psi_x \psi_x) \> \,.
\eeq
Thus the insertion of the EOM operator
simply multiplies the observable with the number of propagators $n_p$ needed to compute it.
Thus the leading contribution of $x$ and $y$ being close together in the
vector four-point function ($V_\mu\equiv \bar\psi\gamma_\mu \psi$) is
\beq\eqlab{2v_final}
\int \frac{d\Omega_k}{2\pi^2}
\Big\< \int d^4x \int d^4y \;e^{ik(x-y)}\, V_\mu(x) V_\mu(y)   V_\sigma(z) V_\lambda(0)\Big\>
\stackrel{k^2\to\infty}{=} \Big( - \frac{6}{k^2} m \frac{\partial}{\partial m} + \frac{4}{k^2}  \Big) \<V_\sigma(z) V_\lambda(0)\>.
\eeq

We now move to the case of $x$ and $y$ simultaneously being in close vicinity of a third current at position $z$.
In terms of Wick contractions, the relevant case is where the connecting point is $z$,
i.e.\ there are propagators $(y\to z \to x)$ or $(x\to z \to y)$.
The cases where the connecting point is $x$ or $y$ do not contribute to the O($1/k^2$) behaviour.
A straightforward if somewhat tedious calculation then gives
\bea\eqlab{3v_final}
    && \int \frac{d\Omega_k}{2\pi^2}\int d^4x \int d^4y \; e^{ik(x-y)} \;
    \bar\psi_x \gamma_\mu \psi_x\; \bar\psi_z \gamma_\sigma \psi_z\; \bar\psi_y \gamma_\mu \psi_y
    \stackrel{k^2\to\infty}{=}  -\frac{2}{k^2}    \bar\psi_z \gamma_\sigma \psi_z  .
    \eea
    The same contribution appears when $(x,y)$ are close to the origin, thus doubling this contribution
    in the four-point function of the vector current.

Altogether, 
from Eqs.\ \eref{2v_final} and \eref{3v_final}, we then find in leading order of the OPE
    \bea\eqlab{OPELO}
&&  \int\frac{d\Omega_k}{2\pi^2}  \Big\< \int d^4x \int d^4y \;e^{ik(x-y)}\, V_\mu(x) V_\mu(y)  V_\sigma(z) V_\lambda(0)\Big\>
     \stackrel{k^2\to\infty}{=}
    - \frac{6}{k^2} m \frac{\partial}{\partial m} \<V_\sigma(z) V_\lambda(0)\>.
    \eea
The terms not leading to mass-derivatives cancel, and only the mass-derivative of the vector two-point function determines the large-$k^2$ asymptotics of the forward LbL amplitude. Thus we have reproduced the very first term in \Eqref{OPE1}.
Acting on both sides of \Eqref{OPELO} with \Eqref{getHVP}
leads to the first term in \Eqref{OPEfinal} (with $n_f=1$ and  ${\cal Q}_f=1$).

The leading-order calculation above is equally valid for the QED as for the QCD four-point function.
In the pure QED context, 
 one easily verifies with the help of Eqs.~\eref{QEDct} and \eref{dm} that the textbook O($e^2$) mass counterterm removes the $\log(\Lambda)$ term
 in $\overline\Pi_{{\rm 4pt}}$ predicted by the OPE.
We have thus verified \Eqref{OPEfinal} in the pure QED case: given that $\overline\Pi_{e^2}(Q^2)$ given in \Eqref{Pi1loop} only depends on $Q^2/m^2$, the effect of inserting the trace anomaly, proportional to $(2Q^2\frac{\partial}{\partial Q^2} + m\frac{\partial}{\partial m})$, would cancel.

\subsection{Cottingham-like formula for the isovector contribution to HVP}

Starting from \Eqref{OPE1} with 
\beq\eqlab{V3}
V^{(1)}_\rho = V^{(2)}_\rho := \frac{1}{2}
(\bar u \gamma_\rho u  - \bar d \gamma_\rho d )
\eeq
the isovector component of the e.m.\ current,
the same steps lead to the analogue of \Eqref{OPEfinal},
with $\overline\Pi_{e^2}$ now replaced by the isovector contribution $\overline\Pi_{e^2}^{33}(Q^2)$ to the leading HVP, and $\overline\Pi_{{\rm 4pt}}^{\gamma\gamma 33}(Q^2,\Lambda)$ the corresponding QED correction to that contribution.
The same steps once again could be taken with the charged isovector currents
\beq\eqlab{V+}
V^{(1)}_\rho :=\frac{1}{\sqrt{2}}\;
\bar u \gamma_\rho d,
\qquad V^{(2)}_\rho :=\frac{1}{\sqrt{2}}\;
\bar d \gamma_\rho u ,
\eeq
which leads to the quantities $\overline\Pi_{e^2}^{-+}(Q^2)$ and its QED correction  $\overline\Pi_{{\rm 4pt}}^{\gamma\gamma-+}(Q^2,\Lambda)$.
Taking the difference of \Eqref{OPE1} obtained once with the choice of currents from \Eqref{V3} and once with the choice of \Eqref{V+}, one finds that all terms on the right-hand side cancel. 
This is clear for the contribution from the insertion of isoscalar operators, whose correlation function is identical with two members of the same isospin multiplet.
For the isovector mass insertion $(\bar u u - \bar d d)$,
$G$-parity ensures that this insertion vanishes separately in the neutral and in the charged isovector channel.
In other words, all counterterms from the action cancel in this difference\footnote{A very similar observation was already made in the case of the pion e.m.\ mass splitting in Ref.~\cite{deDivitiis:2013xla}.}, as has been noted in Ref.~\cite{Bruno:2018ono}, which contains an exploratory lattice-QCD calculation of this quantity. 
However, care must be taken of the fact that, unlike the cases considered so far, the currents of \Eqref{V+} are not gauge invariant with respect to QED. Therefore this case requires further study. We note that, in the radiative corrections to the leptonic decay of a charged pion~\cite{Carrasco:2015xwa}, the e.m.\ correction to a charged-current correlator represents one of several contributions.

Other flavor cases may be of interest, in particular the correlator between the isovector and the isoscalar components of the photon, which vanishes in isospin-symmetric QCD in the absence of quark electric charges. In this case, the action counterterms do not vanish altogether, though only the isovector mass insertion $(\bar u u - \bar d d)$ contributes in the analogue of \Eqref{OPEfinal}.

%% file: hvpnlopi0sec.tex
\section{The $\pi^0$-exchange contribution}\label{sect:pi0}

\begin{figure}[hb]
\centering
	\includegraphics[width=0.6\textwidth]{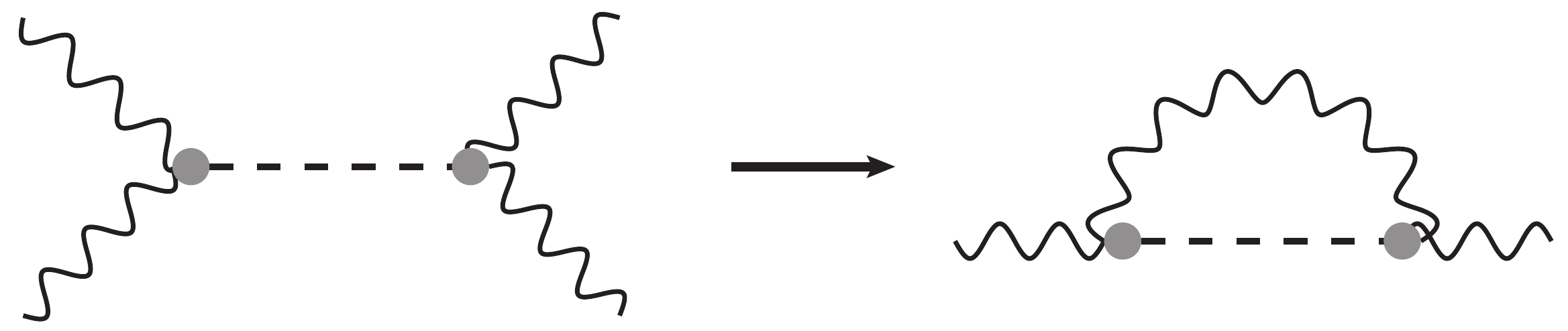}
\caption{$\pi^0$-exchange contribution to the VP via Cottingham-like formula.}
	\label{fig:PionContibution}
\end{figure}

In this section, our goal is to present the form of the $\pi^0$-exchange contribution to the forward HLbL amplitude $\MM$, cf.\ \Figref{PionContibution}, since it is the longest-range contribution.
As in the previous section, we work in Euclidean space.
We define the Fourier transform of the four-point function of this current as in \Eqref{4ptEucl} and the forward amplitude is obtained as in \Eqref{forwEucl}.
The O($e^4$) contribution to the polarization tensor with a  regularized internal photon propagator then reads
\bea\label{eq:poltensE}
\Pi_{{\rm 4pt};\mu\lambda}(q,\Lambda) = -\frac{e^4}{2}\int \frac{d^4k}{(2\pi)^4}\, \Big[ \frac{1}{k^2}\Big]_\Lambda
\;\Pi_{\mu\sigma\sigma\lambda}(q,k,-k)
= (q_\mu q_\lambda- \delta_{\mu\lambda} q^2)\;
\Pi_{{\rm 4pt}}(q^2,\Lambda).
\eea

For the $\pi^0$-exchange contribution, proportional to the square of its transition form factor ${\cal F}$, we have 
\bea
\Pi_{\mu\sigma\sigma\lambda}(q,k,-k)
&=& -\epsilon_{\mu\sigma\alpha\beta}\; \epsilon_{\sigma\lambda\gamma\delta} \;q_{\alpha} \;k_{\beta} \;k_{\gamma} \;q_{\delta}
\;{\cal F}(-q^2,-k^2)^2
\\ && \times \left[ \frac{1}{(q+k)^2+m_\pi^2} + \frac{1}{(q-k)^2+m_\pi^2}\right].
\nonumber\eea
We  perform the angular integration by using the Gegenbauer polynomial expansion of propagators (see for instance~\cite{Knecht:2001qf}), and 
the final expression is
\bea\label{eq:pi0vp}
 \Pi_{{\rm 4pt}}(q^2,\Lambda) &=& \frac{-e^4}{16\pi^2|q|} 
\int_0^\infty d|k|\,|k|^4 \Big[ \frac{1}{k^2} \Big]_\Lambda \,
{\cal F}(-q^2,-k^2)^2 \, Z_{|q|,|k|}^{m_\pi}
\Big( 1 -  {\txts\frac{1}{3}} (Z_{|q|,|k|}^{m_\pi})^2 \Big),
\\ 
Z_{|q|,|k|}^m &=& \frac{1}{2|q| |k|}\Big(q^2+k^2+m^2 - \sqrt{(q^2+k^2+m^2)^2 - 4q^2 k^2}\Big).
\eea
This expression, once inserted into \Eqref{g2VP}, can be viewed as the VP analogue of the Jegerlehner-Nyffeler relation for the $\pi^0$ contribution to HLbL scattering in the muon $(g-2)$~\cite{Jegerlehner:2009ry,Nyffeler:2016gnb}.
In the present case, the kinematics are simpler, and correspondingly $\Delta  a_\upmu^{\pi^0}$ takes the form of a two- rather than three-dimensional integral for a yet to be specified transition form factor ${\cal F}$.
The unsubtracted VP is UV-finite
for fixed $\Lambda$, while the subtracted 
VP $\overline\Pi_{{\rm 4pt}}(q^2,\Lambda)$ remains UV-finite for $\Lambda\to\infty$, unlike in the full QCD case,
as we have seen in section~\ref{sec:ope}, 
when short-distance contributions from quarks are taken into account.

As an example, for the VMD parameterization of the transition form factor,
\beq
{\cal F}(-q_1^2,-q_2^2) = \frac{{\cal F}(0,0)}{(1+q_1^2/m_V^2)(1+q_2^2/m_V^2)},
\eeq
one obtains, near the chiral limit, the singular behaviour 
\bea
\lim_{\Lambda\to\infty}\frac{\partial\Pi_{{\rm 4pt}}}{\partial Q^2}(Q^2=0,\Lambda)  &=&  \frac{\alpha^2}{6 } \;{\cal F}(0,0)^2\;
\Big[ 5 + \log   \left(\frac{m_V^2}{m_\pi^2}\right) +{\rm O}(m_\pi^2/m_V^2)\Big].
 \eea
 For a pion mass which is still heavy relative to the muon mass, 
 the  contribution reads
$ \Delta a_\upmu \simeq \frac{\alpha m_\mu^2}{3 \pi} \;\Pi_{{\rm 4pt}}'(0) $.
 Parametrically, this contribution behaves similarly to the $\pi^0$ contribution
 to the HLbL contribution to $a_\upmu$~\cite{Knecht:2001qg,Blokland:2001pb}, except that in the latter case the chiral logarithm  enters quadratically.

Numerically, with ${\cal F}(0,0)=(4\pi^2 f_\pi)^{-1}$ and $f_\pi=92.4$\,MeV, $m_V=0.77549\,$GeV, and the physical $\pi^0$ mass 
 one obtains
from Eq.~\eqref{eq:pi0vp} with the  QED kernel \eqref{eq:g2VP}
  the following contribution to $a_\upmu$,
 \beq
\Delta a_\upmu^{\pi^0} = 0.370\times 10^{-10}.
 \eeq
 This result agrees  with the value given in~\cite{Blokland:2001pb}.
 We note that the result is more than an order of magnitude smaller than the contribution of the $e^+e^-\to\pi^0\gamma$ channel
 in the dispersive representation of $a_\upmu^{\rm VP}$~\footnote{The vastly different size of the result of Ref.\ \cite{Blokland:2001pb}
 as compared to the $e^+e^-\to\pi^0\gamma$ channel contribution was pointed out to one of us in 2016 by Andreas~Nyffeler.
 See also the recent Ref.\ \cite{Crivellin:2022gfu}, appendix~D.},
 however the quantity $\Delta a_\upmu^{\pi^0}$ computed here is not precisely the same.
We finally remark that the master relation 
Eq.~\eqref{eq:pi0vp} applies equally well to the other pseudoscalar mesons, notably the $\eta$ and $\eta'$.

%% file: lattstrat.tex
\section{Electromagnetic correction to the HVP in lattice QCD: a computational strategy \label{sec:lattstrat}}

The general structure of \Eqref{Pie4tot} also applies to the calculation of the isospin-breaking contribution to the leading HVP $\Pi_{e^2}(Q^2)$ (from QED and strong isospin breaking) in lattice regularization, in which case the inverse lattice spacing $1/a$ plays the role of the UV cutoff.
To be more specific, we note that lattice QCD admits $(N_{\rm f}+1)$ bare parameters, i.e., the SU(3) gauge coupling and the $N_{\rm f}$ quark masses, which we assemble into a vector $\vec {\rm b}^{\,\rm lat}$. When correcting the isosymmetric theory for isospin breaking, the bare parameters must be readjusted.
The shifts $\delta{\rm b}^{\rm lat}_i$ in the bare parameters are determined by requiring that the theory with isospin breaking reproduces $(N_{\rm f}+1)$ suitable experimental observables~\cite{deDivitiis:2013xla}; typically, the masses of hadrons which are stable in the absence of weak interactions,
\beq
M_h^{\rm phys} = M_h^{\rm iso} + M^{\rm lat}_{{\rm 4pt},h}(a;0) + \sum_{i=1}^{N_{\rm f}+1} J^{\rm lat}_h{}^{(i)}\;\delta{\rm b}^{\rm lat}_i
\qquad (h=1,\dots,N_{\rm f}+1),
\eeq
where $M_h^{\rm phys}$ is the experimental hadron mass, $M_h^{\rm iso}$ is its value at the chosen expansion point in isosymmetric QCD, $M^{\rm lat}_{{\rm 4pt},h}(a;M_\gamma)$ is the O($e^2$) e.m.\ contribution computed with (in general) a photon mass $M_\gamma$ and  $J^{\rm lat}_h{}^{(i)}\equiv {\partial M_h}/{\partial{\rm b}^{\rm lat}_i}=\<h|O^{(i)}_{\rm lat}|h\>$ is given by the forward matrix element of the operator conjugate to parameter ${\rm b}_i^{\rm lat}$. Thus, given a lattice calculation of $\vec M^{\rm iso}$, $\vec M^{\rm lat}_{{\rm 4pt}}(a;0)$ and the matrix $J^{\rm lat}$, the vector $\delta\vec{\rm b}^{\,\rm lat}$ is obtained by solving a linear system.

We are now in a position to write the lattice-regularization analogue of \Eqref{Pie4tot} for the subtracted HVP as 
\beq \eqlab{Pie4totlat}
\Delta\overline{\Pi}(Q^2) = \lim_{a\to 0} \Big(\overline{\Pi}^{\rm lat}_{\rm 4pt}(Q^2, a;0)+ {\overline\Pi}^{\rm lat}_{\rm ct}(Q^2, a)\Big),
\eeq 
where the counterterm has the form
\beq\eqlab{Pilatct}
\overline{\Pi}^{\rm lat}_{\rm ct}(Q^2, a)=\sum_{i=1}^{N_{\rm f}+1} \delta{\rm b}^{\rm lat}_i\frac{\partial~~}{\partial{\rm b}^{\rm lat}_i}\,\overline{\Pi}_{e^2}(Q^2),
\eeq
and $\overline{\Pi}^{\rm lat}_{\rm 4pt}(Q^2, a;M_\gamma)$ denotes the e.m.\ four-point function contribution to the HVP, computed (in general) with an internal photon of mass $M_\gamma$.
Massive QED has previously been used to control photon zero modes in finite volume, with the physical limit $M_\gamma\to 0$ taken after extrapolating to infinite volume~\cite{Endres:2015gda, Patella:2017fgk, Bussone:2017xkb, Clark:2022wjy}; our approach, by contrast, is to keep $\Lambda=M_\gamma$ fixed and use it for separating long-range contributions from UV-divergent ones.
Based on the identity of \Eqref{gampropsplit}
for the photon propagator with a fixed $\Lambda\sim400\;$MeV, we propose to perform the following decompositions\footnote{\Eqref{Pi4lat2stp} and
\Eqref{M4pt2stp} hold up to corrections suppressed by one or two powers of the lattice spacing.},
\bea
\overline\Pi^{\rm lat}_{\rm 4pt}(Q^2, a; 0) &=&
\overline\Pi_{\rm 4pt}(Q^2, \Lambda=M_\gamma) + \overline\Pi^{\rm lat}_{\rm 4pt}(Q^2,  a; M_\gamma),
\eqlab{Pi4lat2stp}
\\
\vec M^{\rm lat}_{\rm 4pt}(a; 0) &=& 
\vec M_{\rm 4pt}( \Lambda=M_\gamma)
  +\vec M^{\rm lat}_{\rm 4pt}( a; M_\gamma).
  \eqlab{M4pt2stp}
\eea
Up to the subtraction at $Q^2=0$, the function
$\overline\Pi_{\rm 4pt}(Q^2,\Lambda)$ is the same as
in \Eqref{Pie4tot}. Since $\Lambda$ plays the role of
the Pauli-Villars UV regularization scale, the continuum limit $a\to0$ can be taken for this quantity.
Similarly, $\vec M_{\rm 4pt}( \Lambda)$ represents the e.m.\ hadron-mass corrections computed with the Pauli-Villars regulated photon propagator. The continuum limit can also be taken in this case, since the same OPE, as reviewed in section~\ref{sec:ope}, determines the asymptotic behaviour of the forward Compton amplitude on hadron $h$~\cite{{Hill:2016bjv}}.

Given a choice of $M_\gamma$, each term on the right-hand side of \Eqref{Pi4lat2stp} is affected by rather different systematics  on the lattice and is meant to be evaluated separately.
The same observation applies to the two terms on the right-hand side of \Eqref{M4pt2stp}.
The UV-finite part $\overline\Pi_{\rm 4pt}(Q^2,\Lambda)$ of \Eqref{Pi4lat2stp} receives long-distance contributions due to the long-range photon propagator. A coordinate-space representation free of power-law finite-volume effects is presented below in subsection \ref{subsec:ccs}. The result can be compared to an evaluation based on our Cottingham-like formula, \Eqref{Mdisprep} and \Eqref{CottinghamFormulaFinal}.
For the second term of \Eqref{Pi4lat2stp}, one possible expression in the time-momentum representation~\cite{Bernecker:2011gh}  is 
\bea
\overline\Pi^{\rm lat}_{\rm 4pt}(Q^2, a; M_\gamma)
 &=&  -\frac{e^4}{2}\frac{a^{12}}{L^3L_0} \sum_{z_0>0}    \left(z_0^2 - \frac{4}{Q^2}\sin^2\frac{|Q|z_0}{2} \right)
\sum_k     G^{\rm lat}_{\mu\nu}(k)
\\ && \Big\<\sum_{x,y} e^{ik(x-y)} \sum_{\vec z}\;  V^{\rm em}_\sigma(z)V^{\rm em}_\nu(y)V^{\rm em}_\mu(x) V^{\rm em}_\lambda(0)
 +\; {\rm tadpoles}\Big\>,
\nonumber
\eea
where $V^{\rm em}_\nu(y)$ and $V^{\rm em}_\mu(x)$ are discretized as conserved currents and the simplest form of the photon propagator (in Feynman gauge) is
\beq
G^{\rm lat}_{\mu\nu}(k) =  \frac{\delta_{\mu\nu}}{\hat k^2 + M_\gamma^2}, \qquad
\hat k^2 \equiv {\textstyle\frac{4}{a^2}}\sum_{\mu=0}^3 \sin^2{\textstyle\frac{ak_\mu}{2}},
\eeq
and the tadpole terms ensure the transversality
of the four-point function with respect to contracting it with $\hat k_\mu$ or $\hat k_\nu$.
Similarly, $\vec M^{\rm lat}_{\rm 4pt}(a;M_\gamma)$ can be determined by well-established methods where the (now massive) photon is treated as part of the finite-volume lattice field theory.

We now briefly discuss how to  compute the hadronic mass shifts $\vec M_{\rm 4pt}(\Lambda)$ with a long-range, but Pauli-Villars regulated photon propagator. As for $\overline\Pi_{\rm 4pt}(Q^2,\Lambda)$, it is possible to avoid power-law effects in the volume~\cite{Feng:2018qpx} by using coordinate-space methods
and, additionally, by explicitly correcting the elastic contribution of the forward Compton amplitude for finite-volume effects. As a slight variation to the concrete proposal in~\cite{Feng:2018qpx}, this correction could be done with the help of a separate calculation of the e.m.\ form factor(s) of the hadron whose mass correction is being computed. These methods could also be applied to the difference of $M_{{\rm 4pt},h}(\Lambda)$ between proton and neutron, a quantity that could be compared to predictions based on the original Cottingham formula. 

We remark that the currently most frequently used formulation of
lattice QCD coupled to photons consists in removing the photon
zero-mode in every time-slice~\cite{Hayakawa:2008an}.  The
corresponding finite-size effects on the HVP have recently been
investigated and found to be parametrically of order $1/L^3$, and
numerically small in the framework of scalar
QED~\cite{Bijnens:2019ejw}. Another recent investigation provides a
systematic analysis of various finite-size effects beyond the
pointlike approximation of hadrons, in particular of pseudoscalar
mesons masses~\cite{DiCarlo:2021apt} (see also references therein).
As an alternative method, first results (primarily on hadron masses)
based on simulating QCD+QED with $C^*$ boundary conditions have
recently been presented~\cite{Bushnaq:2022aam}.

In conclusion, while the numerical practicability of the presented method remains to be demonstrated,
we have established that it is possible to avoid power-law finite-volume effects altogether in computing $\Delta{\Pi}(Q^2)$ on the lattice.
How large the discretization errors on this quantity are at finite lattice spacing with the method proposed above will need to be explored in practice. Here we remark that the  Pauli-Villars regularization of the photon propagator is only one of many posssible choices. For instance, with the decomposition
\bea\eqlab{doublePV}
\frac{1}{k^2} &=& \left(\frac{1}{k^2}
- G_{\rm sub}(k^2,\zeta,\Lambda)\right)
+G_{\rm sub}(k^2,\zeta,\Lambda),
\\
G_{\rm sub}(k^2,\zeta,\Lambda) &=& \frac{1}{1 - \zeta}\left(\frac{1}{k^2+\zeta\Lambda^2} -  
\frac{\zeta}{k^2+\Lambda^2} \right),
\qquad 0 < \zeta < 1,
\eea
which amounts to a `double Pauli-Villars' regularization of the photon propagator, the same strategy as described above can be carried out, now with the expression in brackets in \Eqref{doublePV} falling off as fast as $1/k^6$ at large $k^2$.

%% file: coordspace.tex
\subsection{Coordinate-space representation of $\overline\Pi_{{\rm 4pt}}(Q^2,\Lambda) $ free of power-law finite-size effects}
 \label{subsec:ccs}

A Euclidean coordinate-space expression for the subtracted HVP is 
\beq\overline\Pi(Q^2) 
= \int_z H_{\lambda\sigma}(z) \;\tilde \Pi_{\sigma\lambda}(z)\,,
\label{eq:CCS_LO}
\eeq
where the leading contribution is 
$\tilde \Pi_{e^2;\sigma\lambda}(z)=e^2 \,\<V^{\rm em}_\sigma(z)\,V^{\rm em}_\lambda(0)\>$,
the relevant ($Q$-dependent) coordinate-space kernel
$H_{\lambda\sigma}(z)$ was derived in~\cite{Meyer:2017hjv} (Sect.~II.B.2)
and we have abridged $\int_z \equiv \int d^4z$.
Expanding a QCD correlation function to second order in the e.m.\ coupling leads to the insertion of the product of two e.m.\ currents,
whose relative positions are weighted by the internal photon propagator.
Thus, using Feynman gauge for the latter, we arrive at the expression
\beq
 \overline\Pi_{{\rm 4pt}}(Q^2,\Lambda) =  -\frac{e^4}{2}\delta_{\mu\nu}\int_{x,y,z} H_{\lambda\sigma}(z)
  \Big[G_0(y-x) \Big]_\Lambda
\; \Big\<V^{\rm em}_\sigma(z)V^{\rm em}_\nu(y)V^{\rm em}_\mu(x) V^{\rm em}_\lambda(0)\Big\>,
\label{eq:CCSmaster}
\eeq
for the regulated contribution to the subtracted HVP. The Pauli-Villars regulated photon propagator in position space reads 
\beq
\Big[G_0(x) \Big]_\Lambda =  \frac{1}{4\pi^2 x^2} - \frac{\Lambda K_1(\Lambda |x|)}{4\pi^2 |x|},
\eeq
which is only logarithmically divergent for $x^2\to0$.
Here $K_1$ is the modified Bessel function of the second kind.
We note a close analogy of expression (\ref{eq:CCSmaster})
with the master relation used for the HLbL contribution to the muon $(g-2)$ in Refs.~\cite{Chao:2020kwq,Chao:2021tvp,Chao:2022xzg}.
Similarly, $a_\upmu^{\rm VP}$ can be obtained from Eq.~\eqref{eq:CCSmaster} by replacing the kernel $H_{\lambda\sigma}$ by the appropriate one given in Sect.~II.B.3 of Ref.\ \cite{Meyer:2017hjv}.
The main feature of our proposal is that no IR-regularization of the photon propagator is needed. Thus finite-size effects are expected to be on the order of $\exp({-m_\pi L/2})$, as in the case of the HLbL contribution~\cite{Chao:2020kwq}.

\subsection{Computing $a_\upmu^{\rm HVP}$: $\pi^0$-exchange contribution to the coordinate-space integrand}

By Fourier-transforming the polarization tensor associated with Eq.\ \eqref{eq:pi0vp} (see Eq.\ \eqref{eq:poltensE}), one obtains
the O($e^4$) contribution  $\tilde \Pi_{{\rm 4pt};\sigma\lambda}(z,\Lambda)$.
After insertion into Eq.~\eqref{eq:CCS_LO} and contraction of the indices, 
the integrand is a scalar function of $|z|$.
We illustrate the integrand of this last scalar integral in the following.
Alternatively to the proposed position-space approach, one can reach the widely used time-momentum representation (TMR)~\cite{Bernecker:2011gh} by  
Fourier-transforming the polarization tensor at vanishing spatial momentum only with respect to $q_0$.

\begin{figure}[t]
    \centering
    \begin{minipage}{0.45\textwidth}
    \includegraphics[scale=0.35]{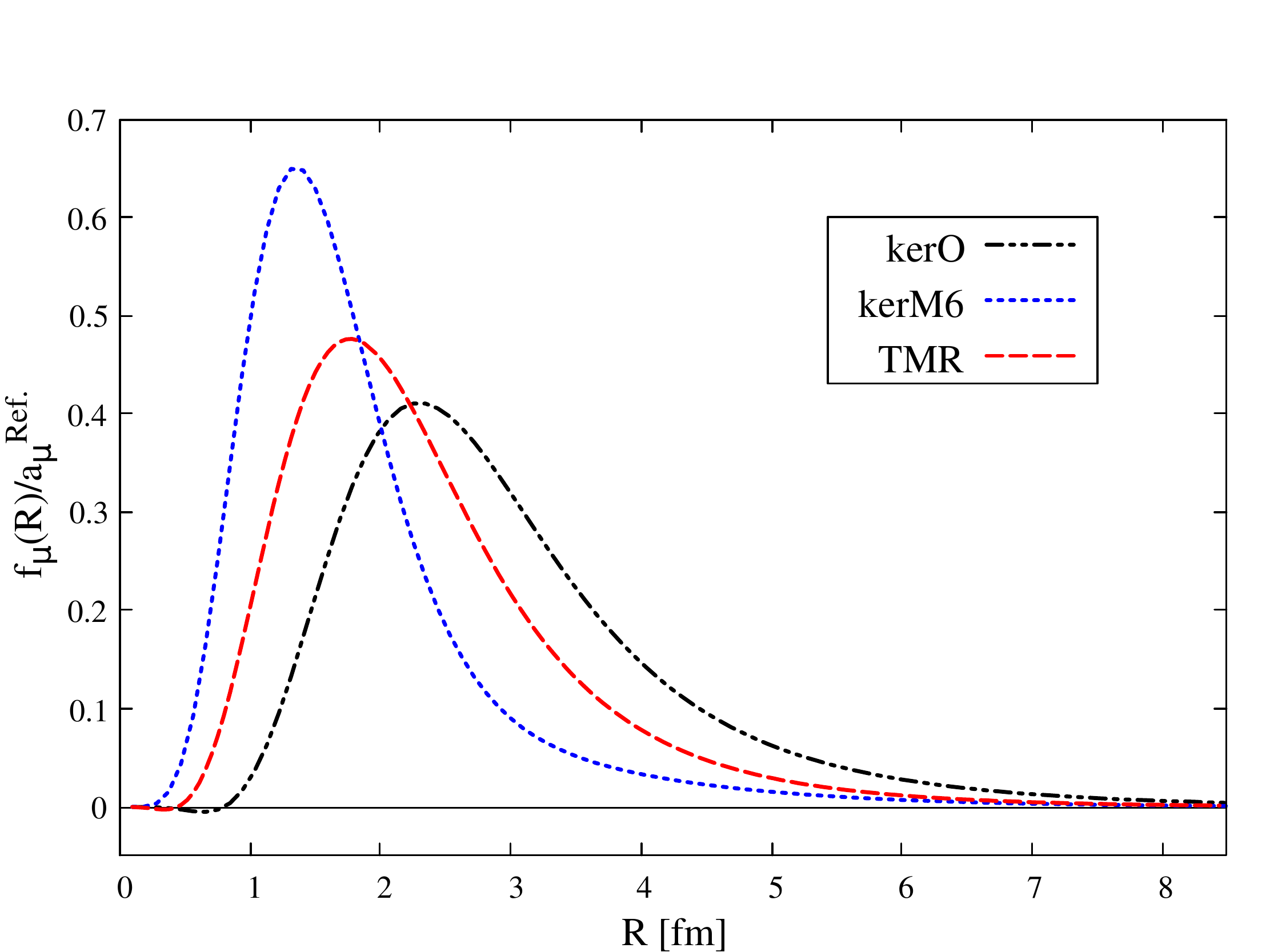}
    \end{minipage}
    \hspace{8pt}
    \begin{minipage}{0.45\textwidth}
    \includegraphics[scale=0.35]{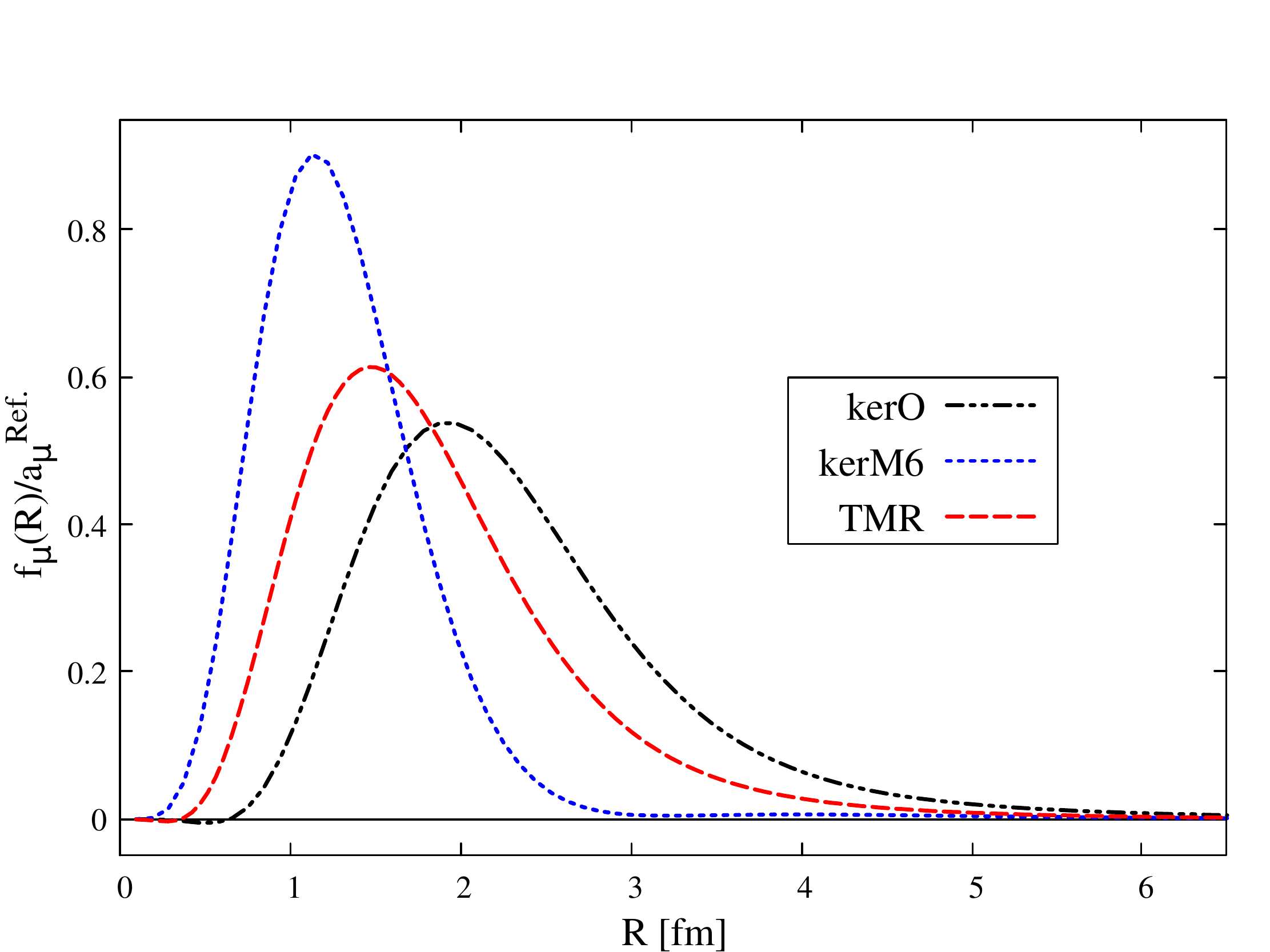}
    \end{minipage}
    \caption{Comparison between the normalized integrands from different representations of the $\pi^0$-exchange contribution to $a_\upmu$ for $\Lambda=3m_\mu$ (left panel) and $\Lambda=200m_\mu$ (right panel), where 'kerO' is obtained with the original coordinate-space kernel, 'kerM6' with improved kernel given in \ref{sect:ccs_kernel}, with the parameter $M$ set to 6 in Eq.~\eqref{eq:fmsub}. The integrands are normalized such that the area under the curves equals  unity.}
    \label{fig:comp_sub}
\end{figure}

Choosing the same parameters as in section~\ref{sect:pi0} for the VMD parameterisation of the transition form factor, we obtain the integrands as functions of $R\equiv|z|$ (and $R\equiv z_0$ for the TMR case) as shown in Fig.~\ref{fig:comp_sub}, where the results with the original~\cite{Meyer:2017hjv} position-space kernel (kerO) and the TMR are compared, calculated 
in the continuum and infinite volume at two different Pauli-Villars masses $\Lambda = 3\,m_\mu$ and $200\,m_\mu$; recall that the $\pi^0$-exchange
contribution by itself remains finite as $\Lambda \to\infty$.
Both integrands are rather long-range, an observation which 
implies a certain difficulty for lattice calculations if the O($e^4$)
contribution is to be computed with good relative precision.
We remark that the integrand displayed in Fig.~\ref{fig:comp_sub} corresponds to the sum of all Wick contractions contributing to the four-point function of the e.m.\ current, but, using the results in appendix~A of Ref.\ \cite{Chao:2021tvp}, it would be fairly straightforward to adapt the prediction to individual Wick contractions of the quark fields.

As a consequence of the Ward-Identity of the vector current, 
a term $\partial_\lambda \left[ z_\sigma F(|z|) \right]$ can be added
to the position-space kernel $H_{\lambda\sigma}(z)$ without 
changing the integrated result of Eq.~\eqref{eq:CCS_LO} in infinite volume~\cite{Ce:2018ziv}.
With a judiciously chosen subtraction, one can make the $z$-integrand in Eq.~\eqref{eq:CCSmaster} more peaked in the small-$|z|$ region.
As in practice, a calculation on the lattice is limited by the degrading signal-to-noise ratio when the arguments of the correlator are far apart in position-space, the possibility of reshaping the integrand makes the position-space representation appealing.
Such a technique has also been used for lattice determinations of the HLbL scattering contribution to
$a_\upmu~$\cite{Blum:2016lnc,Chao:2020kwq}. Figure~\ref{fig:comp_sub} thus also shows the result
of improving the kernel (kerM6) to make the integrand shorter-range; details of its construction are given in~\ref{sect:ccs_kernel}.
For the $\Lambda=3\, m_\mu$ case, the partially-integrated $a_\upmu(R)$ obtained with kerM6 already reaches about 70\% of its final value $a_\upmu(\infty)$ at $R=2$ fm, but only about 50\% with the TMR.
For $\Lambda=200\, m_\mu$, the benefit becomes even more apparent: 95\% with kerM6 and merely about 65\% with the TMR.
We thus expect that the position-space method with an improved kernel should offer a good opportunity to compute the e.m.\ correction to $a_\upmu^{\rm{HVP}}$ on the lattice, with better controlled finite-volume effects.

%% file: PolarizedCrossSections.tex
\section{Applying the Cottingham-like formula to the QED vacuum polarization}
Here we give more details on how the two-loop QED contribution to vacuum polarization is reproduced
via the Cottingham-like formula. We begin with
\Eqref{CottinghamFormulaFinal}, which
by a variable change, $\nu = K Q x$, is cast into:
\bea 
\Pi_{{\rm 4pt}}(Q^2,\Lambda) &=&  \frac{1}{3 (2\pi)^3 Q^2} \int\limits_0^{\infty} d K^2  
\,\int\limits_{0}^{1} d x \, \sqrt{1-x^2}\,
\MM( K Q x ,\,K^2,\,Q^2) ,
\eea 
where $\Lambda$ is the scale regularizing the integral over $K^2$ in the ultraviolet. We keep the regularization implicit throughout this appendix.
Substituting the dispersive representation of the LbL amplitude and the optical theorem, we obtain
 \bea
\Pi_{{\rm 4pt}}(Q^2,\Lambda)
&=& \frac{1}{3(2\pi)^3 Q^2}\int\limits_0^\infty d K^2 \,
\Bigg[ \frac{\pi}{4}
\MM(\bar\nu ,K^2,\,Q^2)\nn\\
  &+& \int\limits_{\nu_\mathrm{thr.}}^\infty d \nu  \left(\frac{2}{\nu+\sqrt{X}} - \frac{\nu}{\nu^{ 2} -\bar{\nu}^2 }\right) \sqrt{X}\,  \sigma(\nu,K^2,\,Q^2) 
 \Bigg],
 \eqlab{DRform}
 \eea
 where $X = \nu^2-Q^2K^2$, and $\bar \nu$ is the subtraction point 
 (below we use $\bar\nu=0$ and $\bar\nu = KQ$
 and verify that the results are equivalent).
 We next provide the ingredients needed to evaluate the vacuum polarization to two-loops in QED, including
 the necessary counterterms. 
 
\subsection{Polarized $\gamma\gamma$ fusion cross sections}
\label{sec:CrossSectionsDef}
 We start with the tree-level cross sections for the QED process $\gamma^\ast\gamma^\ast\to \ell \overline{\ell}$ ($\ell$ denote spinor QED fermions) with polarized virtual photons. The conventions are the same as in Refs.~\cite{Pascalutsa:2012pr} and \cite{Budnev:1975poe}:
\beq
L \equiv \log\left(\frac{1+\sqrt{a}}{\sqrt{1-a}}\right),\quad 
a \equiv \frac{X}{\nu^2}\left(1-\frac{4m^2}{s}\right), \quad X = \nu^2-Q^2K^2,
\eeq
where $s=(k+q)^2=2\nu-K^2-Q^2$, $\nu=k\cdot q$; $Q^2=-q^2$ and $K^2=-k^2$ are the spacelike photon virtualities. The threshold energy in this case is given by $\nu_\mathrm{thr.} = 2m^2 + \nicefrac12 (K^2 +Q^2)$.
The cross sections corresponding to the
fusion of two polarized photons, either transverse ($T$) or longitudinal ($L$), read:
\bea
\sigma_{TT}(\nu,Q^2,K^2)&=&  \frac{1}{2}\left(\sigma_\parallel+\sigma_\perp\right)\nn\\
&=& \frac{\alpha^2\pi}{2}\frac{s^2\nu^3}{X^3}
\Bigg\{
\sqrt{a}\bigg[-4\left(1-\frac{X}{s\nu}\right)^2-(1-a)+\frac{K^2Q^2}{\nu^2}\left(2-\frac{1}{(1-a)}\frac{4X^2}{s^2\nu^2}\right)\bigg]\nn\\
&&
+\bigg[3-a^2+2\left(1-\frac{2X}{s\nu}\right)^2-\frac{2K^2Q^2}{\nu^2}(1+a)\bigg]L
\Bigg\}.
\eea
\bea
\sigma_{LT}(\nu,Q^2,K^2) &=& \sigma_{TL}(\nu,K^2,Q^2) \nn\\
&=& \alpha^2\pi Q^2\frac{s}{\nu X^2}
\Bigg\{
\bigg[\left(\nu-K^2\right)^2\bigg(-2(1-a)-(3-a)\frac{Q^2K^2}{X}\bigg)+2\nu K^2(1+a)
\nn\\
&&-K^4(3+a)\bigg]L+\sqrt{a}\bigg[\left(\nu-K^2\right)^2\left(2+\frac{3Q^2K^2}{X}\right)-2\nu K^2+K^4\frac{3-a}{1-a}\bigg]
\Bigg\}.\qquad\;
\eea
\bea
\sigma_{LL}(\nu,Q^2,K^2)&=&2\alpha^2\pi Q^2K^2\frac{s^2}{\nu X^2}
\Bigg\{
\sqrt{a}\bigg[-2-\frac{3-2a}{1-a}\frac{Q^2K^2}{X}\bigg]\left(2+\frac{3Q^2K^2}{X}\right)L
\Bigg\}.
\eea

The optical theorem connects the absorptive cross sections with the imaginary part of the corresponding forward LbL amplitudes,
\bea
\im \mathcal{M}_{PP'} &=& 2\sqrt{X}\,\sigma_{PP'}, \qquad
P,P' \in\{L,T\}.
\eea

%% file: apdxQED2.tex
The total cross section requied in the Cottingham formula is given by\footnote{Hereafter, until the end of \ref{sec:1loopLbLsubfunc}, we  set $m=1$, without loss of generality.}
\bea
\sigma 
&=& 4\sigma_{TT} - 2 \sigma_{LT} - 2\sigma_{TL} + \sigma_{LL} = \frac{8\pi\alpha^2}{\nu}
 \Bigg\{2\nu\sqrt{X\frac{4+K^2+Q^2-2\nu}{K^2+Q^2-2\nu}}\nn\\
 &\times&\frac{2\nu[K^2Q^2+2+\nu]-K^2[K^2(Q^2-1)+2(1+\nu)]-Q^2[Q^2(K^2-1)+2(1+\nu)]}{K^2Q^2(4-2\nu+K^2+Q^2)-4\nu^2}\nn\\
 &+&
 \left[(K^2+Q^2-\nu)^2+\nu^2+4(\nu-1)\right]
 \log
 \frac{1+\sqrt{\frac{X}{\nu^2}\frac{4+K^2+Q^2-2\nu}{K^2+Q^2-2\nu}}}{\sqrt{1-\frac{X}{\nu^2}\frac{4+K^2+Q^2-2\nu}{K^2+Q^2-2\nu}}}\Bigg\}.
 \eqlab{imMMqed}
 \eea

\subsection{The one-loop LbL amplitudes
and subtraction functions } \label{sec:1loopLbLsubfunc}
The forward one-loop LbL amplitude has been derived  with the help of \textit{Package-X} \cite{Patel:2015tea,Patel:2016fam} and is given by:
 \bea
 &&\MM(\nu, K^2,Q^2) \nn\\&&
 = 16\alpha^2\Bigg(
 6-\Bigg\{
\frac{2 \log \left[\frac{1}{2} Q \left(\sqrt{Q^2+4}+Q\right)+1\right]}{\sqrt{Q^2+4}}\nn\\&&
\times \bigg(-4 \nu ^2 Q^2 \left[\left(K^2-2\right) \left(K^2+1\right)
   Q^4+\left(K^2+2\right) \left(7 K^2-2\right) Q^2+6 K^4+52 K^2+16\right]\nn\\&&
   +K^2 Q^4 \left(K^2+Q^2+4\right)^2 \left[K^2 \left(Q^2+4\right)-2
   Q^2+4\right]+96 \nu ^4\bigg)
   \bigg/ 
   \bigg(
   K^4 Q^5
   \left(K^2+Q^2+4\right)^2\nn\\&&
   +16 \nu ^4 Q
   -4 K^2 \nu ^2 Q^3 \left[K^2 \left(Q^2+2\right)+2 \left(Q^2+4\right)\right]
   \bigg)+\big\{K\leftrightarrow Q\big\}
 \Bigg\}
 \nn\\&&
 +
 \Bigg\{
 \frac{2 \sqrt{1+\frac{4}{K^2+2 \nu +Q^2}} \log \left[\frac{1}{2} \left(\sqrt{\left(K^2+2 \nu +Q^2\right) \left(K^2+2 \nu +Q^2+4\right)}+K^2+2 \nu
   +Q^2+2\right)\right]}{K^2 Q^2 \left(K^2+Q^2+2 \nu +4\right)-4 \nu ^2}\nn\\&&
   \times\bigg(K^2Q^2(K^2+Q^2+2\nu)-2(K^2+Q^2)(\nu-1)-(K^4+Q^4)-2\nu(\nu+2)\bigg)\nn\\&&
   +\frac{(K^2+Q^2)^2 +2 \nu (K^2+Q^2) + 2 \nu(\nu-2) -4 }{\nu }\,C_0\left(-K^2,-Q^2,-K^2-2 \nu -Q^2;1,1,1\right)\nn\\&&
   +\big\{\nu\to -\nu\big\}
 \Bigg\}
 \Bigg),
 \eea
 where  $C_0(p_1^2,p_2^2,(p_1+p_2)^2; m_1^2,m_2^2,m_3^2)$ is the scalar one-loop integral in the \textit{LoopTools} \cite{Hahn:1998yk} notation.
 The expressions for the subtraction function for the cases of $\bar\nu=0$ and $\bar\nu = KQ$ are, respectively:
\bea
   \MM(0 ,K^2,\,Q^2) &=& -
   32 \alpha^2 \Bigg\{
   \frac{ 
   Q^6+Q^4\left(3 K ^2+2\right)+2 Q^2\left(K ^4+7 K ^2-4\right)+4K^2(2K^2+5)}{K ^2
   Q \sqrt{Q^2+4} \left(K ^2+Q^2+4\right)}\nn\\
   &&\times\log \left[\frac{1}{2} Q \left(\sqrt{Q^2+4}+Q\right)+1\right]\nn\\
   &&+\frac{K ^6+K ^4 \left(3 Q^2+2\right)+2 K ^2 \left(Q^4+7 Q^2-4\right)+4 Q^2 \left(2
   Q^2+5\right)}{Q^2 K  \sqrt{K ^2+4}  \left(K ^2+Q^2+4\right)}\nn\\
   &&\times\log \left[\frac{1}{2} K  \left(\sqrt{K ^2+4}+K \right)+1\right]\nn\\
   &&-\frac{Q^6+K^6-2(Q^4+K^4)+5Q^2K^2(Q^2+K^2)}{K ^2 Q^2
   \sqrt{\left(K ^2+Q^2\right) \left(K ^2+Q^2+4\right)}}\nn\\
   &&\times\log \left[\frac{1}{2} \left(K ^2+\sqrt{\left(K ^2+Q^2\right) \left(K ^2+Q^2+4\right)}+Q^2+2\right)\right]\nn\\
   &&-2 \left(K ^2+Q^2-2\right) C_0\left(-Q^2,-K ^2,-K ^2-Q^2;1,1,1\right)-3
   \Bigg\},
\eea
\bea
  \MM(KQ ,K^2,\,Q^2)&=& -16 \alpha^2 \Bigg\{-6+\frac{\left(K^2+Q^2-2\right) \left[(K+Q)^2+4\right]^{3/2}}{K^2 Q^2 (K+Q)}\nn\\
  &&\times\log \left[\frac{1}{2} (K+Q)
   \left(\sqrt{(K+Q)^2+4}+K+Q\right)+1\right]\nn\\
  &&+\frac{\left(K^2+Q^2-2\right) \left[(K-Q)^2+4\right]^{3/2} }{K^2 Q^2 | K-Q| }\nn\\
  &&\times\log \left[\frac{1}{2} | K-Q|  \left(| K-Q|
   +\sqrt{(K-Q)^2+4}\right)+1\right]\nn\\
   &&-\frac{2 \left[\left(K^4+2 K^2+28\right) Q^2+K^6+6 K^4-32\right] }{K Q^2 \sqrt{K^2+4}  \left(K^2-Q^2\right)}\nn\\
   &&\times\log \left[\frac{1}{2} K
   \left(\sqrt{K^2+4}+K\right)+1\right]\nn\\
   &&-\frac{2 \left[ \left(Q^4+2 Q^2+28\right)K^2+Q^6+6
   Q^4-32\right] }{K^2 Q \sqrt{Q^2+4}
   \left(Q^2-K^2\right)}\nn\\
   &&\times\log \left[\frac{1}{2} Q \left(\sqrt{Q^2+4}+Q\right)+1\right]
   \Bigg\}.
\eea

The vacuum polarization at $Q^2=0$, needed for renormalization, is, in general, given by:
\beq
\Pi_{{\rm 4pt}}(0,\Lambda) =  \frac{1}{3(2\pi)^3}\int\limits_0^\infty d K^2 \int_0^1 d x\sqrt{1-x^2} \,\frac{\MM(KQx,K^2,Q^2)}{Q^2}\Big|_{Q^2=0}.
\eeq
In the case of two-loop QED, we obtain:
\bea
\Pi_{{\rm 4pt}}(0,\Lambda) &=& -\frac{\alpha^2}{4\pi^2}\,\int\limits_0^\infty d K^2
\left\{
\frac{12  \left(K ^2+2\right) \log \left[\frac{1}{2} K  \left(\sqrt{K
   ^2+4}+K \right)+1\right]}{K ^3 \left(K ^2+4\right)^{5/2}}\right.\nn\\
   &&\left.-\frac{ \left(K ^2+2\right)^2+8}{K ^2 \left(K ^2+4\right)^{2}}\right\},\eqlab{Pi0subtraction}
\eea
which is finite upon any regularization set by $\Lambda$.

%% file: QEDcounterterm.tex
\subsection{QED counterterm in the Pauli-Villars regularization\label{sec:qedctpv}}

The counterterm can be obtained in the following way
\beq\eqlab{QEDct}
\overline{\Pi}_{\mathrm{ct}}(q^2,\Lambda) = -\left[m\delta_2-\delta_m\right]\frac{\partial}{\partial m}\overline{\Pi}_{e^2}(q^2),
\eeq
where in the on-shell renormalization scheme\footnote{See Ref.\ \cite{Peskin:1995ev}, whose notation we borrow, in particular their Eqs.\ (7.28) and (7.91).} and in Minkowski-space notation
\bea \eqlab{dm}
m\delta_2-\delta_m &=& \Sigma_2(p\!\!\!/=m) = \delta m= 3m\frac{\alpha}{2\pi}\left[\frac{1}{4}+\log\frac{\Lambda}{m}\right],\\
\eqlab{Pi1loop}
\overline{\Pi}_{e^2}(q^2) &=& \frac{-2\alpha}{\pi}\int_0^1 d x(1-x)x\log\frac{m^2}{m^2-x(1-x)q^2},
\eea
leading to Eq.~(\ref{NonlocalCounterterm}).

%% file: ccs_kernel.tex
\section{Properties of the coordinate-space kernel for $a_\upmu$}\label{sect:ccs_kernel}
\subsection{A modified kernel based on the $\rho$-meson exchange}
In the coordinate-space formulation of Eq.\ (\ref{eq:CCS_LO}), as a consequence of the conservation of the e.m.\ current one can subtract a total derivative~\cite{Ce:2018ziv}
\begin{equation}
    \partial_\lambda \left[ z_\sigma F(|z|) \right] = \delta_{\lambda\sigma} F(|z|) + \frac{z_\lambda z_\sigma}{|z|}F^\prime(|z|)\,,
\end{equation}
from the original~\cite{Meyer:2017hjv} QED kernel $H_{\lambda\sigma}(z)$, without changing the final integrated result.
The smooth function $F$ is arbitrary, as long as it does not generate boundary terms upon integrating by parts.
More precisely, the kernel $H_{\lambda\sigma}(z)$ has the following structure
\begin{equation}
H_{\lambda\sigma}(z) = -\delta_{\lambda\sigma}\mathcal{H}_1(|z|) + \frac{z_\lambda z_\sigma}{|z|^2}\mathcal{H}_2(|z|)\,.
\end{equation}
After the subtraction, the functions $\mathcal{H}_i$, henceforth referred to as \textit{form factors}, are modified to
\begin{equation}
\bar{\mathcal{H}}_1 (r) = \mathcal{H}_1(r) - F(r) \,,
\quad
\bar{\mathcal{H}}_2(r) = \mathcal{H}_2(r) + r F^\prime (r)\,.
\end{equation}
As the coordinate-space formulation for $a_\upmu$ is obtained by a simple substitution of the e.m. kernel appearing in Eqs.\ (\ref{eq:CCS_LO}) and (\ref{eq:CCSmaster}), we will not introduce new notations in the following discussion.
Throughout this appendix, the e.m. kernel and form factors refer implicitly to the ones related to $a_\upmu$.

To avoid the long-distance region where lattice calculations perform less well, we aim at reshaping the integral representation of $a_\upmu$ into a shorter-ranged one by introducing a phyiscally motivated subtraction to the e.m. kernel.
In the leading-order HVP calculation, the contribution of the $\rho$-meson is dominates over a large distance interval.
The simplest way to model the $\rho$-meson is to represent it as a $\delta$-function in the spectral function (see Eq.~(67) of Ref.~\cite{Meyer:2017hjv}).
Based on the behaviour of the correlator at large separations in this simple model, we can choose a one-parameter subtraction function 
\begin{equation}\label{eq:fmsub}
    F^M(r) = \mathcal{H}_1(r)\left( a + \frac{b}{1+m_\mu r} + \frac{c}{1+(m_\mu r)^2}\right)\,,
\end{equation}
where 
\begin{equation}
a = 1\,,\quad b=\frac{12}{5M}\,,\quad 
c = \frac{750 + 1772M}{125M^2}\,.
\end{equation}
The subtraction term Eq.~\eqref{eq:fmsub} preserves the behavior of the kernel near the origin and at long distances.
This subtraction ensures that the $\rho$-meson contribution from our model with a mass of $M$ times that of the muon falls at least faster then $re^{-Mm_\mu r}\log(m_\mu r)$ at large $r$.

\subsection{Approximants for the form factors}
The form factors require evaluating Meijer's $G$-functions, which are in general computationally costly.
Fortunately, as the asymptotic behaviors of the form factors are known, one can efficiently approximate them according to the size of the argument. 
For the typical size of the boxes that we include in lattice calculations to approach the physical point, the maximal separation between the two vector-current insertion is $\sim 6$ fm.
Eventually, one would also want to have good control over the wrap-around effects of light intermediate states due to periodic boundary conditions.
Accounting for these considerations, useful approximants of the form factors should be accurate up to about a distance of $m_\mu r \sim 13$\,.
The approximants that we choose are of the form 
\begin{equation}
\mathcal{H}^{\rm{appr.}}_i(r) = \frac{8\alpha^2}{3m_\mu^2}\bar{f}_i(r)\hat{r}^4\,,\quad 
\textrm{where }\hat{r}\equiv m_\mu r\,.
\end{equation}
For $\hat{r}\leq 2$, we use
\begin{equation}\label{eq:fbarappr1}
\bar{f}_i(r) = \sum_{j=0}^6 a_{j+1}^{(i)}\left(\frac{\hat{r}}{2}\right)^{2j}
+ \sum_{j=0}^5 b_{j+1}^{(i)}\left(\frac{\hat{r}}{2}\right)^{2j+2}\log\left(\frac{\hat{r}}{2}\right)\,,
\end{equation}
and for $2\leq \hat{r} \leq 13$, rational approximations are used:
\begin{equation}\label{eq:fbarappr2}
    \bar{f}_i(r) = \frac{\sum_{j=0}^4 p_{j+1}^{(i)}\hat{r}^j}{1+\sum_{j=1}^4 q_{j+1}^{(i)}\hat{r}^j}\,,
\end{equation}
where the coefficients $a^{(1,2)}_j$, $b^{(1,2)}_j$, $p^{(1,2)}_j$ and $q^{(1,2)}_j$ are tabulated in Tabs.~\ref{tab:a_coeff}--~\ref{tab:q_coeff}.

For completeness, it is convenient to have a similar type of approximant for the derivative of the form factor $\mathcal{H}_1$, as it is required in the implementation of the subtracted kernel introduced in the previous subsection. 
To this end, we consider the approximant
\begin{equation}
  \left[\frac{d}{dr}\left(\hat{r}^{-4}\mathcal{H}_1(r)\right)\right]^{\rm{appr.}} = \frac{8\alpha^2}{3m_\mu^2}\bar{f}_3(r)\,,
\end{equation}
where, for $\hat{r}\leq 2$ 
\begin{equation}\label{eq:fbarappr3}
	\bar{f_3}(r)= \sum_{i=0}^6\left[\left( a^{(3)}_{i+1}  + b^{(3)}_{i+1}\log\left(\frac{\hat{r}}{2}\right)\right)\left(\frac{\hat{r}}{2}\right)^{2i+1}\right]\,,
\end{equation}
and for $2\leq \hat{r} \leq 13$,
\begin{equation}\label{eq:fbarappr4}
	\bar{f}_{3}(r) = \frac{\sum_{i=0}^4 p^{(3)}_{i+1}\hat{r}^i}{1+\sum_{i=1}^4 q^{(3)}_{i+1}\hat{r}^i}\,,
\end{equation}
where the coefficients $a^{(3)}_j$, $b^{(3)}_j$, $p^{(3)}_j$ and $q^{(3)}_j$ are tabulated in Tabs.~\ref{tab:a_coeff}--~\ref{tab:q_coeff}.

In the considered region $\hat{r}\in[0,13]$, these approximations are accurate up to a relative precision of $5\times 10^{-6}$.

\begin{table}[h!]
\centering
 \begin{tabular}{|c|c|c|c|}
 \hline\hline
    $j$ & $a_j^{(1)}$ & $a_j^{(2)}$ & $a_j^{(3)}$ \\
 \hline
    1 & 0.000759549                 & 0.000434028               & 0.0000670547              \\
    2 & $-0.000128258$              & $-0.0000710376$           & $-0.000128956$            \\
    3 & $-0.0000943174$             & $-0.0000656352$           & $-0.0000493386$           \\
    4 & $-0.0000195809$             & $-0.0000147283$           & $-8.51914 \times 10^{-6}$ \\
    5 & $-2.37199\times 10^{-6}$    & $-1.87175\times 10^{-6}$  & $-9.14652 \times 10^{-7}$ \\
    6 & $-1.97309\times 10^{-7}$    & $-1.60997\times 10^{-7}$  & $-6.89663 \times 10^{-8}$ \\
    \cline{4-4}
    7 & $-1.21714\times 10^{-8}$    & $-1.01821\times 10^{-8}$   \\
 \cline{1-3}
 \end{tabular}
 \caption{The coefficients $a^{(i)}_j$ for Eq.~\eqref{eq:fbarappr1} and Eq.~\eqref{eq:fbarappr3}}.\label{tab:a_coeff}
 \end{table}

\begin{table}[h!]
\centering
 \begin{tabular}{|c|c|c|c|}
 \hline\hline
    $j$ & $b^{(1)}_{j}$ & $b^{(2)}_j$ & $b^{(3)}_j$\\
 \hline
    1 & 0.000390625               & 0.000260417               & 0.000390625 \\
    2 & 0.000119358               & 0.0000868056              & 0.000238715 \\
    3 & 0.0000188079              & 0.0000144676              & 0.0000564236 \\
    4 & $1.93762 \times 10^{-6}$  & $1.5501\times 10^{-6}$    & $7.7505 \times 10^{-6}$ \\
    5 & $1.43784 \times 10^{-7}$  & $1.1841\times 10^{-7}$    & $7.1892 \times 10^{-7}$ \\
    6 & $8.12427 \times 10^{-9}$  & $6.84149\times 10^{-9}$   & $4.87456 \times 10^{-8}$ \\
\hline
 \end{tabular}
 \caption{The coefficients $b^{(i)}_j$ for Eq.~\eqref{eq:fbarappr1} and Eq.~\eqref{eq:fbarappr3}}\label{tab:b_coeff}
\end{table}

\begin{table}[h!]
    \centering
    \begin{tabular}{|c|c|c|c|}
    \hline\hline
        $i$ & $p^{(1)}_i$ & $p^{(2)}_i$ & $p^{(3)}_i$ \\ 
        \hline
         1  & 0.000758753               & 0.000428977               & $-0.0000396871$ \\
         2  & 0.00104081                & 0.00071103                & $-0.000390356$ \\
         3  & 0.000126642               & 0.0000476469              & $-3.31526 \times 10^{-6}$ \\
         4  & $-1.2882 \times 10^{-8}$  & $-3.88032\times 10^{-8}$  & $9.09887 \times 10^{-8}$ \\
         5  & $2.70969 \times 10^{-10}$ & $4.99942 \times 10^{-10}$ & $-1.18082 \times 10^{-9}$\\
         \hline
    \end{tabular}
    \caption{The coefficients $p^{(i)}_j$ for Eq.~\eqref{eq:fbarappr2} and Eq.~\eqref{eq:fbarappr4}}
    \label{tab:p_coeff}
\end{table}

\begin{table}[h!]
    \centering
    \begin{tabular}{|c|c|c|c|}
    \hline\hline
        $i$ & $q^{(1)}_i$ & $q^{(2)}_i$ & $q^{(3)}_i$ \\ 
        \hline
         2 & 1.39604    & 1.63492       & 1.43041 \\
         3 & 0.4842     & 0.534896      & 0.518476 \\
         4 & 0.0867795  & 0.0993416     & 0.102334  \\
         5 & 0.00487635 & 0.00444518    & 0.00905763 \\
    \hline
    \end{tabular}
    \caption{The coefficients $q^{(i)}_j$ for Eq.~\eqref{eq:fbarappr2} and Eq.~\eqref{eq:fbarappr4}}
    \label{tab:q_coeff}
\end{table}